\documentclass[aps,pra,twocolumn]{revtex4}
\usepackage{amsmath, amssymb, amscd}
\usepackage{graphicx}
\usepackage{subfigure}
\usepackage{fancybox}
\usepackage{fancyvrb}
\usepackage{xcolor}
\usepackage{wrapfig}
\usepackage{xspace}

\DeclareMathOperator{\id}{id}
\DeclareMathOperator{\lcm}{lcm}
\DeclareMathOperator{\re}{Re}
\DeclareMathOperator{\im}{Im}
\DeclareMathOperator{\wt}{wt}

\DeclareMathOperator{\tr}{Tr}
\DeclareMathOperator{\diag}{diag}

\renewcommand{\vec}[1]{\mathbf{#1}}

\newcounter{inpromtcount}
\newcounter{outpromtcount}
\newcommand{\In}{{\scriptstyle In[\arabic{inpromtcount}]:= \quad \setcounter{outpromtcount}{\value{inpromtcount}}\stepcounter{inpromtcount}}}

\newcommand{\Mathematica}{\textsl{Mathematica}\xspace}

\begin{document}

\title{Bell inequalities, classical cryptography and fractals}

\author{E. Shchukin}
\email{evgeny.shchukin@gmail.com}
\affiliation{Arbeitsgruppe Quantenoptik, Institut f\"ur Physik,
Universit\"at Rostock, D-18051 Rostock, Germany}

\begin{abstract}
The relation between the boolean functions and Bell inequalities for qubits is analyzed. The connection
between the maximal quantum violation of a Bell inequality and the nonlinearity of the corresponding boolean
function is discussed. A visualization scheme of boolean functions is proposed. An attempt to classify Bell
inequalities for qubits is made, a weaker result (classification with respect to Jevons group) is obtained.
The fractal structure of the classification is shown. All constructs are illustrated by \Mathematica code.
\end{abstract}

\maketitle

\section{Introduction}

In their famous paper \cite{pr-47-777} Einstein, Podolsky and Rosen (EPR) suggested a Gedankenexperiment
which, as they believed, must prove the incompleteness of quantum mechanics. An interesting analysis of this
problem was given by Bohr \cite{pr-48-696}. Some progress was achieved by Bell \cite{p-1-195}. He showed that
under assumption of the EPR arguments some inequalities must be fulfilled. If Bohr's arguments are correct
then these inequalities can be violated. It was only in $1980$ when the Bohr's arguments were experimentally
verified \cite{prl-45-617, prl-47-460, prl-49-91, prl-49-1804, n-398-189}. Now the arguments by EPR are
considered to be incorrect, but nevertheless it is the work \cite{pr-47-777} that initiated the discussion of
basics of quantum mechanics.

In this work I analyze the connection between the boolean functions theory and Bell inequalities for
multi-qubit systems (which were obtained in \cite{pra-64-032112}). Surprisingly enough, in many aspects Bell
inequalities theory is analogous to the combinatorial problems of computer logic circuits design developed a
year earlier then the Bell's work \cite{p-1-195} appeared. There is also a relation between Bell inequalities
and applications of boolean functions theory to classical cryptography. For example, classification of Bell
inequalities discussed in \cite{pra-64-032112} is closely connected to the Jevons group, studied in
\cite{umtn-04879-3-T, umtn-04879-4-T}, and the maximal quantum violation of a given Bell inequality is
connected to the nonlinearity of the corresponding boolean function. I made an attempt to clarify these
connections. But there are still many open questions, both combinatorial (like classification of Bell
inequalities with respect to the group $\mathcal{G}_n$) and analytical (like calculation of the maximal
quantum violation $v_f$). Another interesting problem is the connection between the maximal quantum violation
and the uncertainty relation for boolean functions. In this work it is shown that the Bell inequalities whose
maximal quantum violation is the largest (Mermin inequalities), minimize this uncertainty relation.

All mathematical constructions discussed in the text are illustrated with \Mathematica. I choose \Mathematica
since it has an extremely flexible and unified programming language and a very rich set of built-in
mathematical functions. Due to this it is possible to present the algorithms illustrating the discussed
quantities in a very compact form. Using \Mathematica, it is possible to code all the illustrating examples
using high-level constructions, avoiding worrying about low-level programming details which have nothing to
do with the problem under study. I do not pretend to give the most effective \Mathematica code for
calculating different features of boolean functions, my goal is to present a compact and ready-to-use working
code. All the examples can be typed in and run provided that they are entered to \Mathematica in the given
order. I also presented the \textsl{C} source code for the fast Walsh-Hadamard transform and a way to turn it
into an executable program which can be used in \Mathematica.

\section{Boolean functions}

The Bell inequalities for a multi-qubit system, which were obtained in \cite{p-1-195}, are closely connected
with boolean functions theory. There is a natural one-to-one correspondence between the set of all $B_n =
2^{2^n}$ boolean functions of $n$ boolean variables and the set of Bell inequalities for $n$-qubits. In this
section I give a short overview of the notions and results of the boolean functions theory which are needed
for applications to Bell inequalities. The most important notion discussed in this section is the
Walsh-Hadamar transform, which turns out to be the coefficients of the Bell inequality corresponding to a
given boolean function.

Here I also introduce a visualization technique of boolean functions which is useful to graphically represent
different classes of boolean functions. This approach is based on the fact that for all $n$ the number of
boolean functions of $n$ variables is a square (in fact, $B_n = B^2_{n-1}$), so that one can associate the
boolean functions with the cells of a $B_{n-1} \times B_{n-1}$ array. Briefly speaking, this is done
numerating boolean functions with integers, interpreting the vectors of their values as binary decompositions
and then using division modulo $B_{n-1}$. In this section I visualize boolean functions with respect to their
degree and uncertainty. In both cases the pictures show some kind of fractal behavior.

\subsection{Boolean vectors}

Let $\mathbf{Z}_2 = \mathbf{Z}/2\mathbf{Z}$ be the finite field with two elements $\mathbf{Z}_2 = \{0, 1\}$.
The sum and the product of elements $a, b \in \mathbf{Z}_2$ are denoted as $a \oplus b$ and $ab$
respectively. The product of $n$ copies of $\mathbf{Z}_2$ we denote as $V_n = \mathbf{Z}^n_2$ and refer to
its elements as ($n$-dimensional) boolean vectors. It is clear that $|V_n| = 2^n$. The notation $\langle
\vec{x}, \vec{y} \rangle$ is used for the scalar product of two boolean vectors $\vec{x} = (x_1, \ldots,
x_n)$ and $\vec{y} = (y_1, \ldots, y_n)$:
\begin{equation}
    \langle \vec{x}, \vec{y} \rangle = x_1 y_1 \oplus \ldots \oplus x_n y_n.
\end{equation}

There is a natural one-to-one correspondence $b_n$ between the set $V_n$ and the set $\mathcal{B}_n = \{0,
\ldots, 2^n-1\}$:
\begin{equation}\label{eq:bn}
    b_n: V_n \ni \vec{x} = (x_1, \ldots, x_n) \to x = \sum^n_{i=1} x_i 2^{n-i} \in \mathcal{B}_n.
\end{equation}
In other words, the boolean vector $\vec{x}$ corresponds to the integer $x$ whose binary representation is
given by $\vec{x}$. The most significant bit in the binary decomposition of $x$ is the first component of
$\vec{x}$ and the least significant bit is the last component. The correspondence $b_n$ is natural in the
following sense. For an integer $m < n$ the set $V_m$ can be identified with a subset of $V_n$ by padding
$m$-dimensional boolean vectors on the left to extend them to $n$ dimensions. Then the diagram
\begin{equation}
\begin{CD}
    V_m @>b_m>> \mathcal{B}_m \\
    \cap @.     \cap @. \\
    V_n @>b_n>> \mathcal{B}_n
\end{CD}
\end{equation}
is commutative. This means that one can apply $b_n$ to $m$-dimensional boolean vectors with $m < n$. The
correspondence $b_n$ is illustrated by Table \ref{tbl:b}.
\begin{table}[ht]
\begin{tabular}{|c|c|c|c|c|}                                                                 \hline
    $\vec{x}$                  & $b_n(\vec{x})$ & $m=1$    & $m=2$       & $m=3$          \\ \hline
    $(0, 0, \ldots, 0, 0, 0)$  & $0$            & $b_1(0)$ & $b_2(0, 0)$ & $b_3(0, 0, 0)$ \\ \hline
    $(0, 0, \ldots, 0, 0, 1)$  & $1$            & $b_1(1)$ & $b_2(0, 1)$ & $b_3(0, 0, 1)$ \\ \hline
    $(0, 0, \ldots, 0, 1, 0)$  & $2$            &          & $b_2(1, 0)$ & $b_3(0, 1, 0)$ \\ \cline{1-2}\cline{4-5}
    $(0, 0, \ldots, 0, 1, 1)$  & $3$            &          & $b_2(1, 1)$ & $b_3(0, 1, 1)$ \\ \cline{1-2}\cline{4-5}
    $(0, 0, \ldots, 1, 0, 0)$  & $4$            &          &             & $b_3(1, 0, 0)$ \\ \cline{1-2}\cline{5-5}
    \ldots                     & \ldots         &          &             & \ldots         \\ \cline{1-2}\cline{5-5}
    $(0, 0, \ldots, 1, 1, 1)$  & $7$            &          &             & $b_3(1, 1, 1)$ \\ \cline{1-2}\cline{5-5}
    \ldots                     & \ldots         &          &             &                \\ \cline{1-2}
    $(1, 0, \ldots, 0, 0, 0)$  & $2^{n-1}$      &          &             &                \\ \cline{1-2}
    \ldots                     & \ldots         &          &             &                \\ \cline{1-2}
    $(1, 1, \ldots, 1, 1, 1)$  & $2^n-1$        &          &             &                \\ \hline
\end{tabular}
\caption{The correspondence $b_n$ between boolean vectors $\vec{x} \in V_n$ and integer numbers $x \in
\mathcal{B}_n$.} \label{tbl:b}
\end{table}

The set $V_n$ can be ordered in many ways. I use the lexicographical order: for $\vec{x} = (x_1, \ldots,
x_n)$ and $\vec{y} = (y_1, \ldots, y_n)$ the notation $\vec{x} <^{\mathrm{lex}} \vec{y}$ means that there is
$k$, $1 \leqslant k \leqslant n$, such that $x_i = y_i$ for $i = 1, \ldots, k-1$, but $x_k = 0$ and $y_k =
1$. Here $0$ and $1$ are elements of $\mathbf{Z}_2$, but if one considers them to be integers, the same can
be shorter formulated as: $\vec{x} <^{\mathrm{lex}} \vec{y}$ if and only if the first non-zero difference
$y_1 - x_1, \ldots, y_n - x_n$ is positive. Since
\begin{equation}
    \sum^{k-1}_{i=1} 2^i = 2^k-1 < 2^k,
\end{equation}
it is clear that $\vec{x} <^{\mathrm{lex}} \vec{y}$ if and only if $x < y$, where $x = b_n(\vec{x})$ and $y =
b_n(\vec{y})$. This means that $b_n$ preserves the order:
\begin{equation}
    (V_n, <^{\mathrm{lex}}) \simeq (\mathcal{B}_n, <).
\end{equation}

There are two very important functions on $V_n$: Hemming weight $\wt(\vec{x})$ and Hemming distance
$d(\vec{x}, \vec{y})$. The Hemming weight $\wt(\vec{x})$ of a boolean vector $\vec{x} \in V_n$ is defined to
be the number of non-zero components of $\vec{x}$. The Hemming distance between two boolean vectors $\vec{x},
\vec{y} \in V_n$ is the number of positions where components of $\vec{x}$ and $\vec{y}$ differ. It is clear
that the distance can be expressed in terms of the Hemming weight as $d(\vec{x}, \vec{y}) = \wt(\vec{x}
\oplus \vec{y})$. Both the notions play an important role in different applications of boolean functions
theory.

\subsection{Boolean functions}

A boolean function $f(\vec{x}) \equiv f(x_1, \ldots, x_n)$ of $n$ boolean variables is a map $f: V_n \to
\mathbf{Z}_2$. The set of all boolean functions of $n$ boolean variables we denote as $F_n$. There is a
one-to-one correspondence between $F_n$ and $V_{2^n}$:
\begin{equation}\label{eq:FV}
    F_n \ni f \to \bigl(f(b^{-1}_n(2^n-1)), \ldots, f(b^{-1}_n(0))\bigr) \in V_{2^n}.
\end{equation}
Due to this we have $|F_n| = 2^{2^n}$.

Any boolean function $f \in \mathcal{F}_n$ can be represented in the following form:
\begin{equation}\label{eq:mobius}
    f(\vec{x}) = \bigoplus_{\vec{y} \in V_n} g_f(\vec{y}) \vec{x}^\vec{y},
\end{equation}
where $g_f \in F_n$ and $\vec{x}^\vec{y} = x^{y_1}_1 \ldots x^{y_n}_n$ (with $0^0 = 1$) or in the equivalent
form as
\begin{equation}\label{eq:fcoeff}
    f(\vec{x}) = c_0 \oplus \bigoplus^n_{k=1} \bigoplus_{1 \leqslant i_1 < \ldots < i_k \leqslant n}
    c_{i_1 \ldots i_k} x_{i_1} \ldots x_{i_k}.
\end{equation}
The relation \eqref{eq:mobius} can be inverted and the function $g_f$ is expressed through $f$ as
\begin{equation}
    g_f(\vec{y}) = \bigoplus_{\vec{x} \preccurlyeq \vec{y}} f(\vec{x}),
\end{equation}
where $\vec{x} \preccurlyeq \vec{y}$ means that $x_k \leqslant y_k$ for $k = 1, \ldots, n$ (note that
$\preccurlyeq$ is not a linear order). From this it follows that the relation $f \leftrightarrow g_f$ is
one-to-one, which is referred to as M\"{o}bius transform. It is an involution: if $g_f = h$ then $g_h = f$,
or $g_{g_f} = f$ for all $f \in F_n$.

The degree $\deg f$ of a boolean function $f \in F_n$ is defined to be the maximal number $d$ such that there
is $\vec{y} \in V_n$ with $\wt(\vec{y}) = d$ and $g_f(\vec{y})=1$. Boolean functions of degree $1$ are called
affine; they can be represented as follows:
\begin{equation}\label{eq:ba}
    a(\vec{x}) = c_0 \oplus c_1 x_1 \oplus \ldots \oplus c_n x_n
               = c_0 \oplus \langle \vec{c}, \vec{x} \rangle,
\end{equation}
where $c_0 = g_f(0, \ldots, 0)$, $c_i = g_f(0, \ldots, 1, \ldots, 0)$ ($1$ is on the $i$-th position) and
$\vec{c} = (c_1, \ldots, c_n)$. The set of all affine functions is denoted as $A_n$. If $c_0 = 0$ then the
affine function is called linear; such functions are denoted as $l$ (so that $l(\vec{x}) = \langle \vec{c},
\vec{x} \rangle$). The set of all linear functions is denoted as $L_n$. It is clear that $|L_n| = 2^n$ and
$|A_n| = 2^{n+1}$.

A boolean function $f \in F_n$ is called homogeneous of degree $k$ if $g_f(\vec{y}) = 0$ for all $\vec{y} \in
V_n$ with $\wt(\vec{y}) \not= k$. If, in addition, $g_f(\vec{y}) = 1$ for all $\vec{y} \in V_n$ with
$\wt(\vec{y}) = k$, then $f$ is referred to as the $k$-th symmetric function and denoted as $s_k$:
\begin{equation}\label{eq:s}
    s_k(x_1, \ldots, x_n) = \bigoplus_{1 \leqslant i_1 < \ldots < i_k \leqslant n} x_{i_1} \ldots x_{i_k}.
\end{equation}

For boolean functions one can introduce the Hemming weight and distance in the same way as it was done for
boolean vectors: $\wt(f)$ is defined to be the number of $\vec{x} \in V_n$ with $f(\vec{x})=1$, and $d(f, g)$
is the number of $\vec{x} \in V_n$ with $f(\vec{x}) \not= g(\vec{x})$. A boolean function is called
\textit{balanced} if $\wt(f) = 2^{n-1}$. If we denote $f \oplus g \in F_n$ the point-wise sum of $f$ and $g$,
\begin{equation}
    (f \oplus g)(\vec{x}) = f(\vec{x}) \oplus g(\vec{x}),
\end{equation}
then $d(f, g) = \wt(f \oplus g)$. The distance $d(f, M)$ between a boolean function $f \in \mathcal{F}_n$ and
a nonempty subset $M \subseteq \mathcal{F}_n$ is defined via
\begin{equation}
    d(f, M) = \min_{g \in M} d(f, g).
\end{equation}
The distance $N_f$ between $f$ and $A_n$ is called nonlinearity of $f$: $N_f = d(f, A_n)$. It is a very
important characteristic of boolean functions.

\subsection{Walsch-Hadamar transform}

In studying properties of boolean functions the notion of Walsch-Hadamard transform can be very useful. The
Walsch-Hadamard transform of a boolean function $f \in F_n$ is the integer-valued function $W_f$ of $n$
boolean arguments defined via
\begin{equation}\label{eq:WH}
    W_f(\vec{u}) = \sum_{\vec{x} \in V_n} (-1)^{f(\vec{x}) \oplus \langle \vec{x}, \vec{u} \rangle}.
\end{equation}
Let us introduce the Hadamard matrix $\tilde{H}$,
\begin{equation}
    \tilde{H} =
    \begin{pmatrix}
        1 & 1 \\
        1 & -1
    \end{pmatrix}.
\end{equation}
The tensor product $H_n = \tilde{H}^{\otimes n}$ of $n$ copies of $\tilde{H}$ reads as
\begin{equation}
    H_n = ((-1)^{\langle \vec{u}, \vec{v} \rangle})_{\vec{u}, \vec{v} \in V_n}.
\end{equation}
For any $f \in F_n$ we define two boolean vectors $\vec{w}_f, \vec{z}_f \in V_{2^n}$ via
\begin{equation}
    \vec{w}_f = (W_f(\vec{u}))_{\vec{u} \in V_n}, \quad
    \vec{z}_f = ((-1)^{f(\vec{x})})_{\vec{x} \in V_n}.
\end{equation}
The Walsch-Hadamard transform \eqref{eq:WH} can be written in the following compact form:
\begin{equation}\label{eq:WH2}
    \vec{w}_f = H_n \vec{z}_f.
\end{equation}
To calculate any component of $\vec{w}_f$ according to the definition \eqref{eq:WH} it takes $2^n-1$
additions; to calculate all $2^n$ components it takes $2^n(2^n-1) \sim 2^{2n}$additions. Using special
properties of the matrix $H_n$, from \eqref{eq:WH2} it is possible to calculate the vector $\vec{w}_f$ (given
$\vec{z}_f$) with only $n 2^n$ additions (fast Walsch-Hadamard transform). The algorithm and its realization
in $\textsl{C}$ (to be used in \Mathematica) are described in Appendix B.

Note that all components of $\vec{w}_f$ are even, since a sum of an even number of equal terms is even. As an
example, let us calculate $\vec{w}_l$ for a linear function $l(\vec{x}) = \langle \vec{c}, \vec{x} \rangle$.
We need the following relation:
\begin{equation}\label{eq:2n}
    \sum_{\vec{u} \in V_n} (-1)^{\langle \vec{x}, \vec{u} \rangle} = 2^n \delta_{\vec{x}, \vec{0}}.
\end{equation}
This relation is obvious due to the equality
\begin{equation}
    \sum_{\vec{u} \in V_n} (-1)^{\langle \vec{x}, \vec{u} \rangle} = \prod^n_{k=1} (1+(-1)^{x_k}).
\end{equation}
Using the relation \eqref{eq:2n} from the definition \eqref{eq:WH} one can easily get
\begin{equation}\label{eq:l}
    W_l(\vec{u}) = 2^n \delta_{\vec{u}, \vec{c}}, \quad
    \vec{w}_l = 2^n \vec{e}_{\vec{c}},
\end{equation}
where $\vec{e}_{\vec{c}} = (0, \ldots, 1, \ldots, 0)$ ($1$ is on the $b_n(\vec{c})$-th position).

The Walch-Hadamard transform \eqref{eq:WH} is invertible; the inverse transform reads as
\begin{equation}\label{eq:WHinv}
    (-1)^{f(\vec{x})} = \frac{1}{2^n} \sum_{\vec{u} \in V_n} (-1)^{\langle \vec{x}, \vec{u} \rangle}
    W_f(\vec{u}),
\end{equation}
or, in matrix notation, as
\begin{equation}
    \vec{z}_f = 2^{-n} H_n \vec{w}_f.
\end{equation}
This relation is a trivial consequence of \eqref{eq:WH2} due to the simple fact that $H^2_n = 2^n E_{2^n}$,
or $H^{-1}_n = 2^{-n} H_n$. Using the same algorithm of fast Walch-Hadamard transform, one can quickly
calculate $\vec{z}_f$ given $\vec{w}_f$.

Below we will need the following statement: \textit{for an integer-valued function $W(\vec{u})$, $\vec{u} \in
V_n$ there is $f \in F_n$ such that $W_f(\vec{u}) = W(\vec{u})$) if and only if $W(\vec{u})$ satisfies the
condition}
\begin{equation}\label{eq:WHc}
    \sum_{\vec{u} \in V_n} W(\vec{u}) W(\vec{u} \oplus \vec{v}) =
    \begin{cases}
        2^{2n} & \text{if}\quad \vec{v} = \vec{0}, \\
        0 & \text{if}\quad \vec{v} \not= \vec{0}.
    \end{cases}
\end{equation}
\textit{In particular, the Walsch-Hadamar transform $W_f(\vec{u})$ of any boolean function $f \in F_n$
satisfies the Parseval equality:}
\begin{equation}\label{eq:pt}
    \sum_{\vec{u} \in V_n} W^2_f(\vec{u}) = 2^{2n}.
\end{equation}
The proof of this statement is quite simple. The Walsh-Hadamard transform of any boolean function $f \in
\mathcal{F}_n$ satisfies the condition \eqref{eq:WHc}, which can be easily derived from the definition
\eqref{eq:WH} using the equality \eqref{eq:2n}. The less trivial part is to prove that the condition
\eqref{eq:WHc} is sufficient. It is enough to prove that the function
\begin{equation}
    F(\vec{x}) = \frac{1}{2^n} \sum_{\vec{u} \in V_n} (-1)^{\langle \vec{x}, \vec{u} \rangle} W(\vec{u})
\end{equation}
takes only two values $\pm 1$. In other words, it is sufficient to prove that $F^2(\vec{x}) = 1$ for all
$\vec{x} \in V_n$. We have
\begin{equation}
    F^2(\vec{x}) = \frac{1}{2^{2n}} \sum_{\vec{u}, \vec{v} \in V_n} (-1)^{\langle \vec{x}, \vec{u} \rangle \oplus \langle \vec{x}, \vec{v} \rangle} W(\vec{u}) W(\vec{v}).
\end{equation}
For any fixed $\vec{u} \in V_n$ the inner summation over $\vec{v} \in V_n$ is equivalent to the summation
over $\vec{u} \oplus \vec{v}$:
\begin{equation}
    F^2(\vec{x}) = \frac{1}{2^{2n}} \sum_{\vec{u}, \vec{v} \in V_n} (-1)^{\langle \vec{x}, \vec{v} \rangle}
    W(\vec{u}) W(\vec{u} \oplus \vec{v}).
\end{equation}
Changing the summation order we get
\begin{equation}
    F^2(\vec{x}) = \frac{1}{2^{2n}} \sum_{\vec{v} \in V_n} (-1)^{\langle \vec{x}, \vec{v} \rangle} \sum_{\vec{u} \in V_n}
    W(\vec{u}) W(\vec{u} \oplus \vec{v}).
\end{equation}
Due to our assumption the inner sum differs from zero only for $\vec{v} = \vec{0}$ and finally we get
$F^2(\vec{x}) = 1$. This completes the proof.

For nonlinearity one can get the following result:
\begin{equation}\label{eq:Nl}
    N_f = 2^{n-1} - \frac{1}{2} \max_{\vec{u} \in V_n} |W_f(\vec{u})|.
\end{equation}
Let us denote $NW_f = |\{\vec{u} \in V_n | W_f(\vec{u}) \not= 0\}|$, the number of non-zero components of
$\vec{w}_f$. Due to the Parseval equality \eqref{eq:pt} we have
\begin{equation}
    2^{2n} = \sum_{\vec{u} \in V_n} W^2_f(\vec{u}) \leqslant NW_f \max_{\vec{u} \in V_n} |W_f(\vec{u})|^2,
\end{equation}
and from \eqref{eq:Nl} we get the inequality
\begin{equation}\label{eq:N2}
    N_f \leqslant 2^{n-1} - \frac{2^{n-1}}{\sqrt{NW_f}}.
\end{equation}
For example, for a linear function $l \in L_n$ from \eqref{eq:l} we have $NW_f=1$ and from \eqref{eq:N2} it
follows $N_l = 0$.

It is also possible to get an absolute upper bound for nonlinearity. According to the Parseval equality
\eqref{eq:pt} we have
\begin{equation}
    \max_{\vec{u} \in V_n} |W_f(\vec{u})| \geqslant 2^{n/2},
\end{equation}
and due to the relation \eqref{eq:Nl} we get
\begin{equation}
    N_f \leqslant 2^{n-1} - 2^{n/2-1}.
\end{equation}
As we will see, for an even $n$ this bound is exact. For an odd $n$ the exact bound is unknown.

\subsection{Walsh-Hadamard transform as a representation}

Now another point of view on the Walsch-Hadamard transform will be presented. From the definition
\eqref{eq:WH} one can easily derive the following equality:
\begin{equation}\label{eq:Wfg}
    \sum_{\vec{w} \in V_n} W_f(\vec{u} \oplus \vec{w}) W_g(\vec{w} \oplus \vec{v}) =
    2^n W_{f \oplus g}(\vec{u} \oplus \vec{v}),
\end{equation}
valid for all $f, g \in F_n$. In partial case of $f=g$ it reads as
\begin{equation}\label{eq:Wff}
    \sum_{\vec{w} \in V_n} W_f(\vec{u} \oplus \vec{w}) W_f(\vec{w} \oplus \vec{v}) = 2^{2n}
    \delta_{\vec{u}, \vec{v}}.
\end{equation}
Let us for any $f \in F_n$ introduce the $2^n \times 2^n$-matrix $W_f$ via
\begin{equation}
    W_f = \frac{1}{2^n} (W_f(\vec{u} \oplus \vec{v}))_{\vec{u}, \vec{v} \in V_n}.
\end{equation}
The equality \eqref{eq:Wff} means that any matrix $W_f$ is a nontrivial root of the identity matrix: $W^2_f =
E_{2^n}$. The equality \eqref{eq:Wfg} means that for all $f, g \in F_n$
\begin{equation}\label{eq:WW}
    W_f W_g = W_{f \oplus g}.
\end{equation}
In other words, the map $W$,
\begin{equation}
    W: F_n \to \mathrm{GL}(\mathbf{R}^{2^n}), \quad f \to W_f,
\end{equation}
is a representation of the additive group $F_n$ in $\mathbf{R}^{2^n}$. Its character is given by
\begin{equation}
    \chi(f) \equiv \tr(W_f) = (-1)^{\wt(f)}.
\end{equation}

Since $F_n$ is a commutative group, $W$ is equivalent to a direct sum of one-dimensional representations.
This decomposition can be given explicitly. First, let us find eigenvalues an eigenvectors of $W_f$. Namely,
let us show that for all $\vec{w} \in V_n$ the $b_n(\vec{w})$-th column $\vec{z}_{\vec{w}}$ of $H_n$,
\begin{equation}
    \vec{z}(\vec{w}) = ((-1)^{\langle \vec{u}, \vec{w} \rangle})_{\vec{u} \in V_n},
\end{equation}
is an eigenvector of $W_f$ with the corresponding eigenvalue $(-1)^{f(\vec{w})}$:
\begin{equation}\label{eq:Weigen}
\begin{split}
    (W_f &\vec{z}(\vec{w}))_{\vec{u}} =
    \sum_{\vec{v} \in V_n} W_f(\vec{u} \oplus \vec{v}) (-1)^{\langle \vec{v}, \vec{w} \rangle} \\
    &= \sum_{\vec{v} \in V_n} (-1)^{\langle \vec{u} \oplus \vec{v}, \vec{w} \rangle} W_f(\vec{v}) \\
    &= (-1)^{f(\vec{w})} (-1)^{\langle \vec{u}, \vec{w} \rangle}
    = (-1)^{f(\vec{w})} \vec{z}(\vec{w})_{\vec{u}}.
\end{split}
\end{equation}
This means that the matrices $W_f$ are diagonal in the basis $\{\vec{z}(\vec{w})\}_{\vec{w} \in V_n}$. In
particular, for the determinant of $W_f$ we have
\begin{equation}\label{eq:detWf}
    \det W_f = (-1)^{\wt(f)}.
\end{equation}
The diagonalization of $W_f$ reads as
\begin{equation}\label{eq:WHdiag}
    W_f = 2^{-n} H_n \diag((-1)^{f(\vec{x})}_{\vec{x} \in V_n}) H_n.
\end{equation}
For any fixed $\vec{w} \in V_n$ the map
\begin{equation}\label{eq:frep}
    f \to (-1)^{f(\vec{w})}
\end{equation}
is a one-dimensional representation of $F_n$ and the equality \eqref{eq:WHdiag} gives the decomposition of
$W$ into the direct sum of such representations. The line spanned by $\vec{z}(\vec{w})$ is an invariant
subspace of $\mathbf{R}^{2^n}$ on which $W$ acts as \eqref{eq:frep}.

\subsection{Autocorrelation function}

Another important notion is the autocorrelation function $\Delta_{f, g}(\vec{u})$ of two boolean functions
$f, g \in F_n$, which is defined as follows:
\begin{equation}\label{eq:Delta}
\begin{split}
    \Delta_{f, g}(\vec{u}) &= \sum_{\vec{x} \in V_n} (-1)^{f(\vec{x}) \oplus g(\vec{x} \oplus \vec{u})} \\
    &= \frac{1}{2^n} \sum_{\vec{v} \in V_n} (-1)^{\langle \vec{u}, \vec{v} \rangle} W_f(\vec{v}) W_g(\vec{v}).
\end{split}
\end{equation}
Since the map $\vec{x} \to \vec{x} \oplus \vec{u}$ is one-to-one for a fixed $\vec{u}$ and idempotent, i.e.
$(\vec{x} \oplus \vec{u}) \oplus \vec{u} = \vec{x}$ for all $\vec{x}$ and $\vec{u}$, the autocorrelation
$\Delta_{f, g}(\vec{u})$ is symmetric with respect to $f$ and $g$: $\Delta_{f, g}(\vec{u}) = \Delta_{g,
f}(\vec{u})$. For $f = g$ the autocorrelation $\Delta_{f, f}(\vec{u}) \equiv \Delta_f(\vec{u})$ is referred
to as the autocorrelation of the function $f$.

\subsection{Physical interpretation}

\begin{figure}
    \includegraphics{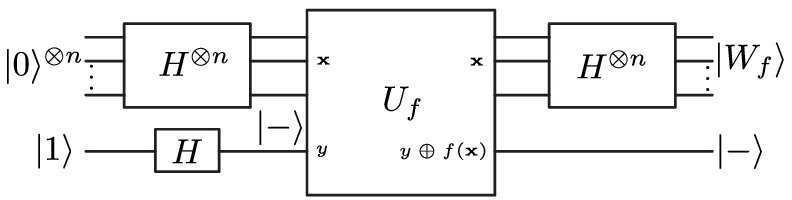}
\caption{Figure}\label{fig:W}
\end{figure}

Here another interpretation of Walsch-Hadamard transform is presented. Consider an $n$-qubit system. For any
$f \in F_n$ let us introduce the state $|W_f\rangle$ via
\begin{equation}
    |W_f\rangle = \frac{1}{2^n} \sum_{\vec{u} \in V_n} W_f(\vec{u}) |\vec{u}\rangle.
\end{equation}
The state $|W_f\rangle$ can be obtained using the circuit shown in Fig.~\ref{fig:W}. Here the Hadamard gate
$H$ is given by the matrix $H = 2^{-1/2} \tilde{H}$ and the $(n+1)$-gubit gate $U_f$ acts as
\begin{equation}
    U_f(|\vec{x}\rangle|y\rangle) = |\vec{x}\rangle|y \oplus f(\vec{x})\rangle.
\end{equation}
Straightforward algebraical manipulations show that if the input state is $|0\rangle^{\otimes n}|1\rangle$
then the output state is $|W_f\rangle|-\rangle$, where
\begin{equation}
    |\pm\rangle = \frac{1}{\sqrt{2}} (|0\rangle \pm |1\rangle).
\end{equation}

Due to the Parseval equality \eqref{eq:pt} the states $|W_f\rangle$ are normalized:
\begin{equation}
    \langle W_f|W_f\rangle = \frac{1}{2^{2n}} \sum_{\vec{u} \in V_n} W^2_f(\vec{u}) = 1,
\end{equation}
but they are not orthogonal:
\begin{equation}\label{eq:WfWg}
    \langle W_f|W_g\rangle = \frac{1}{2^{2n}} \sum_{\vec{u} \in V_n} W_f(\vec{u}) W_g(\vec{u})
                           = \frac{1}{2^n} \Delta_{f, g}(\vec{0}).
\end{equation}
From this we can conclude: $|W_f\rangle$ and $|W_g\rangle$ are orthogonal if and only if $\Delta_{f,
g}(\vec{0}) = 0$.

For any single-qubit operator $A$ and for any $\vec{v} = (v_1, \ldots, v_n) \in V_n$ we use the notation
\begin{equation}
    A^{\vec{v}} = A^{v_1} \otimes \ldots \otimes A^{v_n}
\end{equation}
for the $n$-qubit factorizable gate, with $k$-th component to be $A$ (if $v_k=1$) or $1$ (if $v_k=0$). The
operator with the matrix
\begin{equation}
    X =
    \begin{pmatrix}
        0 & 1 \\
        1 & 0
    \end{pmatrix}
\end{equation}
is called \textsl{NOT} gate; it acts as $X|0\rangle = |1\rangle$, $X|1\rangle = |0\rangle$. It is clear that
for any $\vec{v} \in V_n$
\begin{equation}
    X^{\vec{v}} |\vec{u}\rangle = |\vec{u} \oplus \vec{v}\rangle.
\end{equation}
From the equality \eqref{eq:WHc} we have
\begin{equation}
    \langle W_f|X^{\vec{v}}|W_f\rangle = \frac{1}{2^{2n}} \sum_{\vec{u} \in V_n}
    W_f(\vec{u}) W_f(\vec{u} \oplus \vec{v}) = \delta_{\vec{v}, \vec{0}},
\end{equation}
i.e. all diagonal matrix elements of the from $\langle W_f|X^{\vec{v}}|W_f\rangle$ of any non-trivial
operator (with $\vec{v} \not= \vec{0}$) operator $X^{\vec{v}}$ are equal to zero.

Let us also consider the phase gate
\begin{equation}
    Z =
    \begin{pmatrix}
        1 & 0 \\
        0 & -1
    \end{pmatrix}.
\end{equation}
It acts as $Z|0\rangle = |0\rangle$, $Z|1\rangle = -|1\rangle$. It is clear that
\begin{equation}
    Z^{\vec{v}} |\vec{u}\rangle = (-1)^{\langle \vec{u}, \vec{v} \rangle} |\vec{u}\rangle.
\end{equation}
There is a generalization of the equality \eqref{eq:WfWg}:
\begin{equation}
    \langle W_f | Z^{\vec{v}} | W_g \rangle = \frac{1}{2^n} \Delta_{f, g}(\vec{v}),
\end{equation}
valid for all $\vec{v} \in V_n$.

For any state $|W_f\rangle$ let us consider $2^n$ states
\begin{equation}
    X^{\vec{v}} |W_f\rangle = \frac{1}{2^n} \sum_{\vec{u} \in V_n} W_f(\vec{u} \oplus \vec{v}) |\vec{u}\rangle.
\end{equation}
Below it will be shown that
\begin{equation}
    X^{\vec{v}} |W_f\rangle = |W_{f_\vec{v}}\rangle,
\end{equation}
where $f_\vec{v}(\vec{x}) = f(\vec{x}) \oplus \langle \vec{x}, \vec{v} \rangle$. Due to the relation
\eqref{eq:WfWg} the states $|W_{f_\vec{v}}\rangle$ and $|W_{f_\vec{w}}\rangle$ with different $\vec{v}$ and
$\vec{w}$ are orthogonal:
\begin{equation}
    \langle W_{f_{\vec{v}}}|W_{f_{\vec{w}}}\rangle = \delta_{\vec{v}, \vec{w}}.
\end{equation}

One can easily prove that for all $\vec{v} \in V_n$
\begin{equation}\label{eq:Wuv}
    \sum_{\vec{u} \in V_n} W_f(\vec{u} \oplus \vec{v}) = \sum_{\vec{u} \in V_n} W_f(\vec{u}) =
    (-1)^{f(\vec{0})} 2^n,
\end{equation}
from which we get the following relation
\begin{equation}\label{eq:Wfv}
    \sum_{\vec{v} \in V_n} |W_{f_{\vec{v}}}\rangle = (-1)^{f(\vec{0})} \sum_{\vec{u} \in V_n}
    |\vec{u}\rangle.
\end{equation}
Since the matrix $W_f$ is non-degenerate according to \eqref{eq:detWf}, for any fixed $f \in \mathcal{F}_n$
the $2^n$ vectors $|W_{f_\vec{v}}\rangle$, $\vec{v} \in V_n$ form a basis of the state space of $n$ qubits.
The equality \eqref{eq:Wfv} shows that the sum of the new basic vectors (the diagonal of the parallelepiped
spanned by the new basic vectors) is modulo the sign equal to the old one.

\subsection{Uncertainty relation}

Let $N\Delta_f$ be the number of $\vec{u} \in V_n$ with $\Delta_f(\vec{u}) \not= 0$. Then the following
statement is valid \cite{boolean-functions}: \textit{for all $f \in F_n$ the numbers $NW_f$ and $N\Delta_f$
satisfy the inequality}
\begin{equation}\label{eq:ur}
    NW_f N\Delta_f \geqslant 2^n.
\end{equation}
The quantity $U(f) \geqslant 1$, defined via
\begin{equation}\label{eq:U}
    U(f) = \frac{1}{2^n} NW_f N\Delta_f,
\end{equation}
is referred to as the uncertainty of $f \in F_n$.

In cryptographical applications of boolean functions theory the numbers $NW_f$ and $N\Delta_f$ play an
important role. In some applications it is necessary to use boolean functions $f$ with small $NW_f$, in
others with small $N\Delta_f$. The inequality \eqref{eq:ur} shows that both the numbers $NW_f$ and
$N\Delta_f$ cannot be small, they are subject to \eqref{eq:ur}. In this sense the inequality \eqref{eq:ur}
can be called the uncertainty relation for boolean functions.

\subsection{Visualization of boolean functions}

Now an approach to the visualization of the set of boolean functions will be presented. The set $F_n$ can be
identified with the set $\mathcal{B}_{2^n}$ using the map $b_n$ (remember \eqref{eq:bn} and \eqref{eq:FV}):
\begin{equation}
    B_n=b_{2^n}: F_n \ni f \to \sum_{\vec{x} \in V_n} f(\vec{x}) 2^{b_n(\vec{x})} \in \mathcal{B}_{2^n}.
\end{equation}
Then with each function $f \in F_n$ one can associate a pair of integers $(i, j)$ such that
\begin{equation}
    B_n(f) = (i-1) 2^{2^{n-1}} + (j-1).
\end{equation}
When $f$ varies over $\mathcal{F}_n$, the corresponding pair $(i, j)$ runs over the square
\begin{equation}
    \mathrm{Sq}_n = \{(i, j)| 1 \leqslant i, j \leqslant 2^{2^{n-1}}\}.
\end{equation}
As an example, the correspondence between $f \in F_2$ and $\mathrm{Sq}_2$ is shown in the Table \ref{tbl:b2}.
This square can be depicted as a $2^{2^{n-1}} \times 2^{2^{n-1}}$ array of cells, with each cell marked with
a definite color. For example, one can mark the cells corresponding to the functions of the same degree with
the same color. Below we present pictures of $F_n$ colored with respect to different characteristics of
boolean functions, not only degree.

\begin{table}
\begin{tabular}{|c|c|c|c|c|c|c|c|} \hline
    $f(x_1, x_2)$ & $(0, 0)$ & $(0, 1)$ & $(1, 0)$ & $(1, 1)$ & $B_2(f)$ & $(i, j)$ \\ \hline
    $0$ & $0$      & $0$      & $0$      & $0$      & $1$         & $(1, 1)$ \\ \hline
    $(1+x_1)(1+x_2)$  & $1$      & $0$      & $0$      & $0$      & $2$         & $(1, 2)$ \\ \hline
    $x_2 + x_1 x_2$ & $0$      & $1$      & $0$      & $0$      & $3$         & $(1, 3)$ \\ \hline
    $x_2$   & $1$      & $1$      & $0$      & $0$      & $4$         & $(1, 4)$ \\ \hline
    $x_1 + x_1 x_2$  & $0$      & $0$      & $1$      & $0$      & $5$         & $(2, 1)$ \\ \hline
    $x_1$   & $1$      & $0$      & $1$      & $0$      & $6$         & $(2, 2)$ \\ \hline
    $x_1 + x_2$   & $0$      & $1$      & $1$      & $0$      & $7$         & $(2, 3)$ \\ \hline
    $1+x_1 x_2$   & $1$      & $1$      & $1$      & $0$      & $8$         & $(2, 4)$ \\ \hline
    $x_1 x_2$ & $0$      & $0$      & $0$      & $1$      & $9$         & $(3, 1)$ \\ \hline
    $1+ x_1 + x_2$   & $1$      & $0$      & $0$      & $1$      & $10$         & $(3, 2)$ \\ \hline
    $1+x_1$   & $0$      & $1$      & $0$      & $1$      & $11$         & $(3, 3)$ \\ \hline
    $1+x_1 + x_1 x_2$   & $1$      & $1$      & $0$      & $1$      & $12$         & $(3, 4)$ \\ \hline
    $1+x_2$    & $0$      & $0$      & $1$      & $1$      & $13$         & $(4, 1)$ \\ \hline
    $1+x_2 + x_1 x_2$  & $1$      & $0$      & $1$      & $1$      & $14$         & $(4, 2)$ \\ \hline
    $x_1 + x_2 +x_1 x_2$ & $0$      & $1$      & $1$      & $1$      & $15$         & $(4, 3)$ \\ \hline
    $1$ & $1$      & $1$      & $1$      & $1$      & $16$         & $(4, 4)$ \\ \hline
\end{tabular}
\caption{Boolean function visualization for $n=2$.} \label{tbl:b2}
\end{table}

\begin{figure*}
    \subfigure[$n=2$]{\includegraphics[scale=0.9]{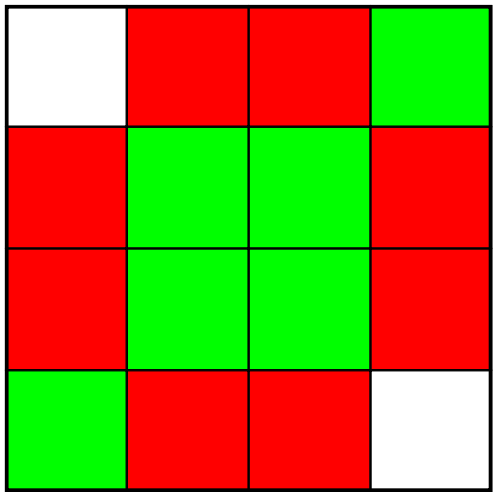}} \hspace*{5mm}
    \subfigure[$n=3$]{\includegraphics[scale=0.9]{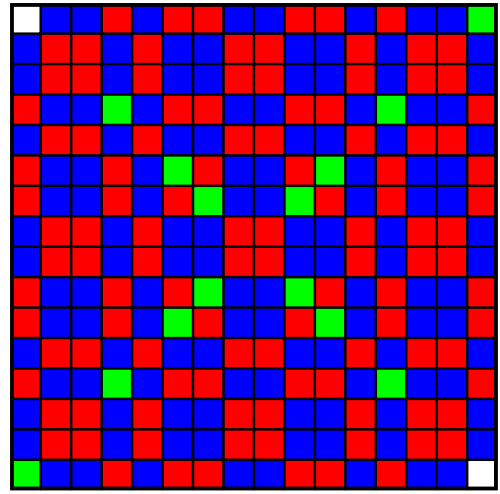}} \\
    \subfigure[$n=4$]{\includegraphics[scale=1.8]{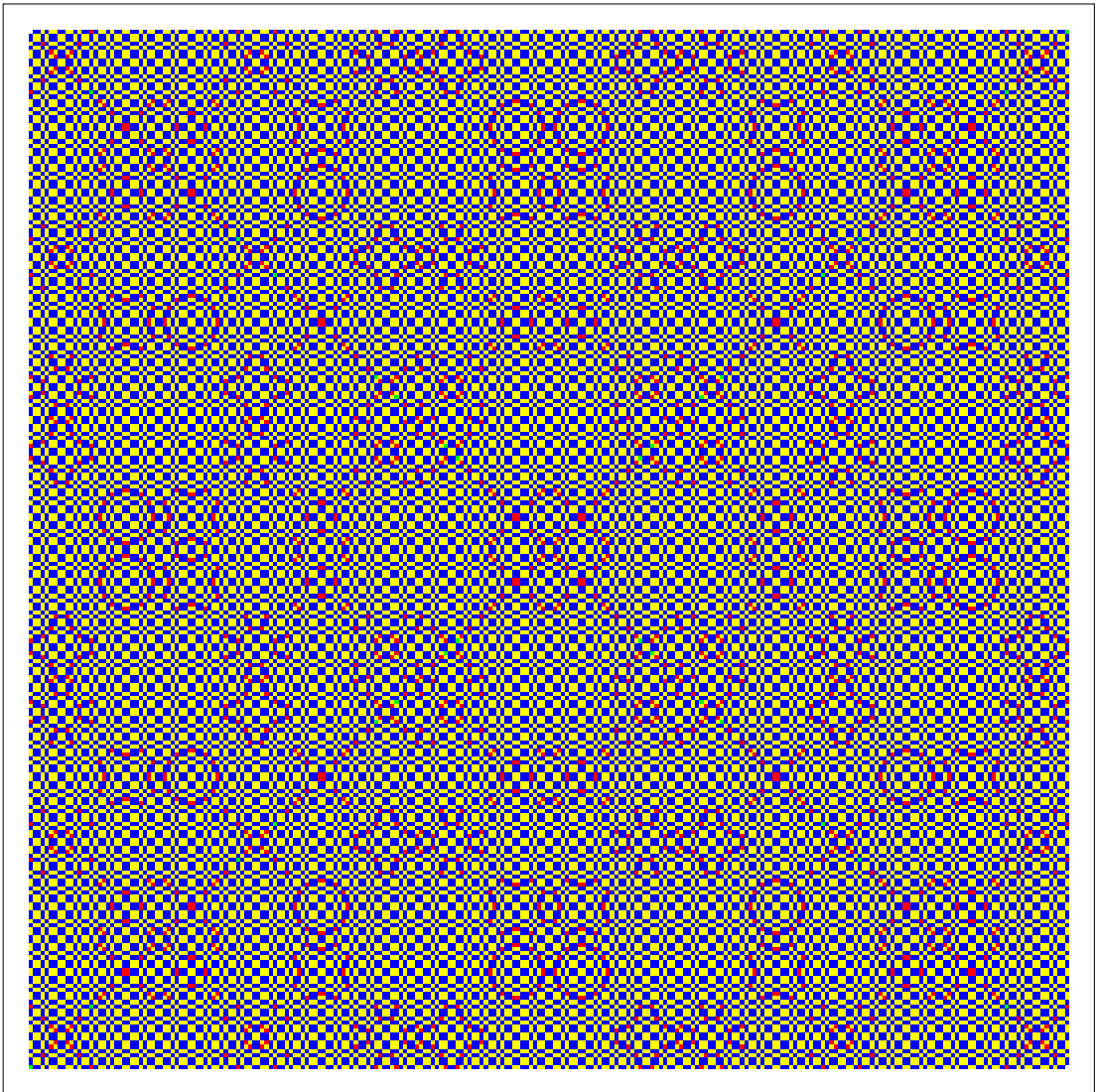}}
\caption{Boolean functions of different orders. White corresponds to the zero order (constant) functions,
green to the first order, red to the second order, yellow to the third order and blue to the forth order
functions.} \label{fig:o}
\end{figure*}

\begin{figure*}
    \subfigure[$n=2$. White corresponds to $U(f)=1$.]{\includegraphics[scale=0.9]{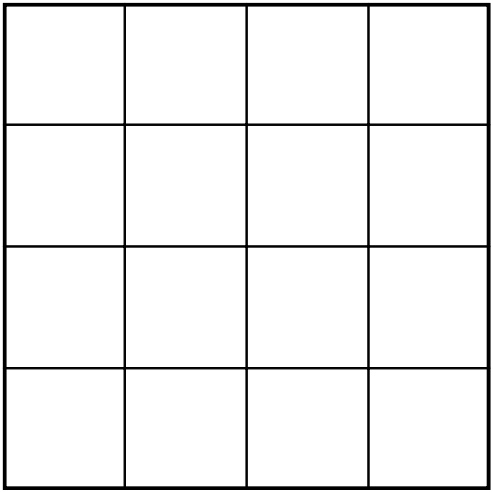}} \hspace*{5mm}
    \subfigure[$n=3$. Light gray corresponds to $U(f)=1$, black to $8$.]{\includegraphics[scale=0.9]{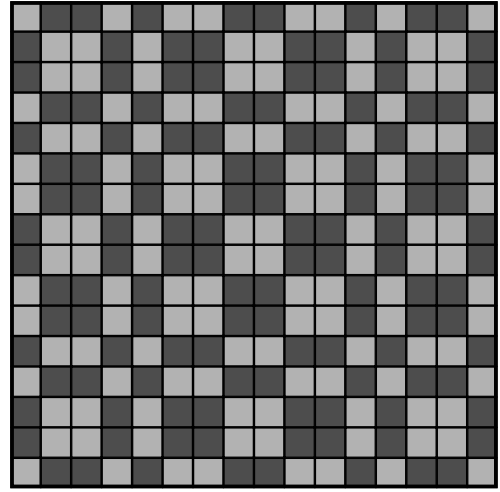}} \\
    \subfigure[$n=4$. White corresponds to $U(f)=1$, yellow to $35/8$, red to $8$ and blue to $16$.]{\includegraphics[scale=1.8]{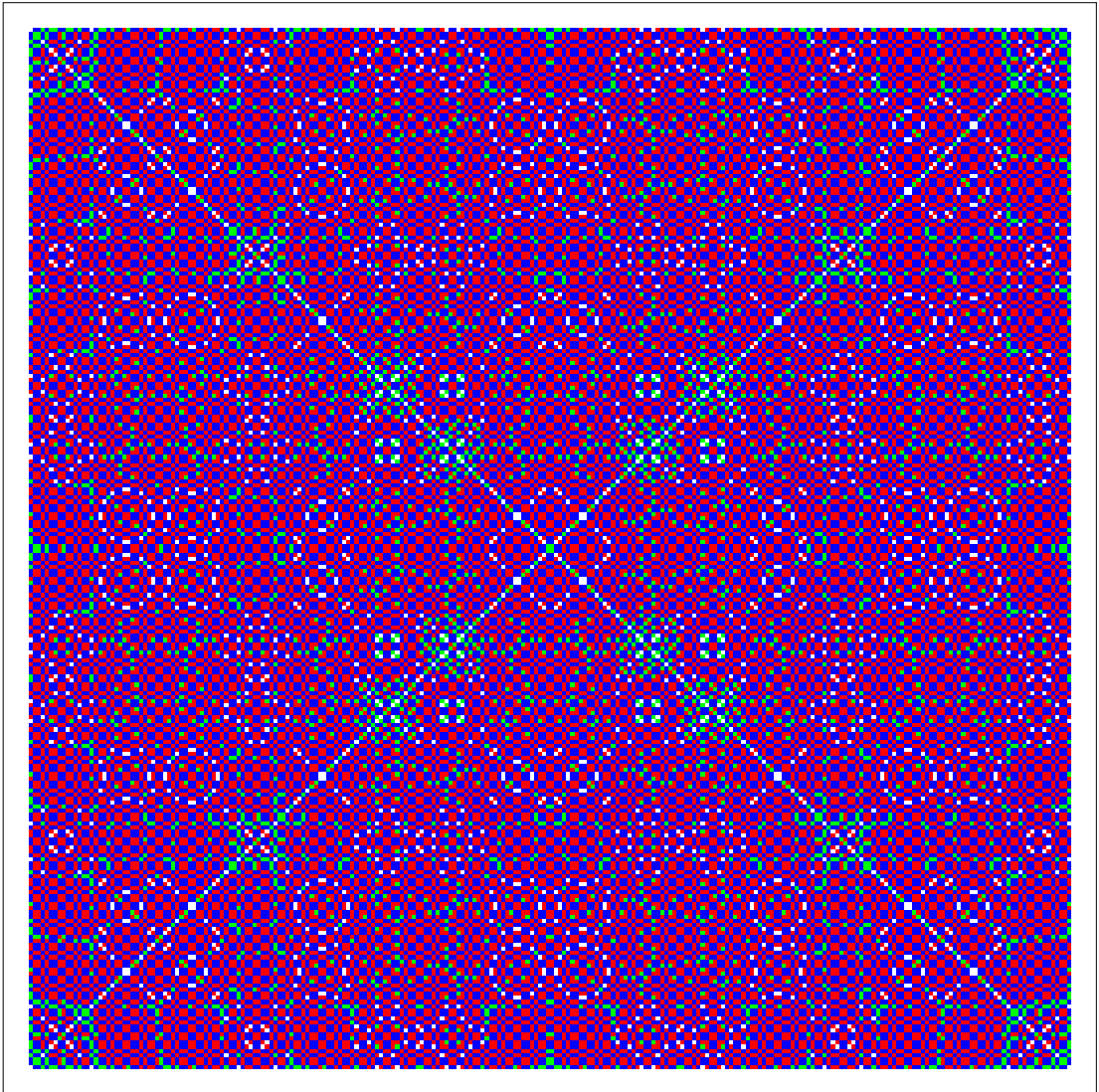}}
\caption{Uncertainties for three different cases.} \label{fig:u}
\end{figure*}

Let us start with the visualization of boolean functions of different orders. In Fig.~\ref{fig:o} the cases
of $n=2$, $n=3$ and $n=4$ are shown. Each subfigure contains $n+1$ different colors. Then let us visualize
boolean functions in the three cases with respect to the uncertainty $U(f)$, \eqref{eq:U}. Fig.~\ref{fig:u}
shows the same three cases of $n=2$, $n=3$ and $n=4$.

\section{Bell inequalities}

In this section the Bell inequalities for $n$ qubits and especially the extremal subclass of Mermin
inequalities are considered and the relation between the maximal violation of a given Bell inequality and the
nonlinearity of the corresponding boolean function is discussed.

\subsection{Construction of Bell inequalities}

Consider $n$ pairs of random variables $A_i(0)$, $A_i(1)$, $i = 1, \ldots, n$ taking only two values $\pm 1$.
Let $E(\vec{u})$, $\vec{u} = (u_1, \ldots, u_n) \in V_n$ be the mathematical expectation of the product of
$n$ variables $A_i(u_i)$, $i = 1, \ldots, n$:
\begin{equation}\label{eq:E}
    E(\vec{u}) = \mathbf{M}(A_1(u_1) \ldots A_n(u_n)).
\end{equation}
Clearly, all $2^n$ expectations $E(\vec{u})$, $\vec{u} \in V_n$, satisfy the inequality
\begin{equation}\label{eq:E1}
    |E(\vec{u})| \leqslant 1.
\end{equation}
Can any $2^n$ numbers $E(\vec{u})$ subject to \eqref{eq:E1} be the mathematical expectations according to
\eqref{eq:E}? The answer is negative.

Let us illustrate this statement by a simple example in the case of $n=2$. For $E(u_1, u_2)$ we have
\begin{equation}
    E(u_1, u_2) = 2\mathbf{P}(A_1(u_1) = A_2(u_2)) - 1.
\end{equation}
From this expression one can conclude that
\begin{equation} \label{eq:E+1}
    E(u_1, u_2) = 1 \Leftrightarrow \mathbf{P}(A_1(u_1) = A_2(u_2)) = 1,
\end{equation}
and that
\begin{equation} \label{eq:E-1}
    E(u_1, u_2) = -1 \Leftrightarrow \mathbf{P}(A_1(u_1) = A_2(u_2)) = 0.
\end{equation}
It is easy to see that the numbers $(1, 1, 1, -1)$ cannot be the expectations $(E(0, 0), E(0, 1), E(1, 0),
E(1, 1))$. In fact, from \eqref{eq:E+1} for $(u_1, u_2) = (0, 0)$, $(0, 1)$ and $(1, 0)$ it follows that in
such a case for any realization of $A_1(0)$, $A_1(1)$, $A_2(0)$ and $A_2(1)$ we would have $A_1(1) = A_2(0) =
A_1(0) = A_2(1)$, but from \eqref{eq:E-1} for $(u_1, u_2) = (1, 1)$ we would have $A_1(1) \not= A_2(1)$. This
contradiction proves that the numbers $(1, 1, 1, -1)$ cannot be the expectations \eqref{eq:E}.

Now a set of necessary and sufficient conditions for $E(\vec{u})$, $\vec{u} \in V_n$ to be the expectations
\eqref{eq:E} is derived. These conditions were obtained in \cite{pra-64-032112}. Our approach follows the
idea of \cite{quant-ph}. For a fixed $f \in F_n$ consider the random variable $A_f$ defined via
\begin{equation}\label{eq:Af}
    A_f = \sum_{\vec{x} \in V_n} (-1)^{f(\vec{x})} \prod^n_{i=1} (A_i(0) + (-1)^{x_i}A_i(1)).
\end{equation}
For any realization of $A_i(u_i)$ the product in \eqref{eq:Af} differs from zero only for one $\vec{x} \in
V_n$, and for this $\vec{x}$ the product is equal to $\pm 2^n$; for all other $2^n-1$ boolean vectors
$\vec{x} \in V_n$ it equals to zero. We see that for any $f \in F_n$ only one term in the sum \eqref{eq:Af}
differs from zero and equals to $\pm 2^n$ so that the sum takes only two values $\pm 2^n$. From this one can
conclude that
\begin{equation}\label{eq:MAf}
    |\mathbf{M}(A_f)| \leqslant 2^n.
\end{equation}
The random variable $A_f$ can be written in the following form:
\begin{equation}
\begin{split}
    A_f &= \sum_{\vec{x} \in V_n} (-1)^{f(\vec{x})} \sum_{\vec{u} \in V_n} (-1)^{(\vec{x}, \vec{u})} A_1(u_1) \ldots A_n(u_n) \\
    &= \sum_{\vec{u} \in V_n} W_f(\vec{u}) A_1(u_1) \ldots A_n(u_n).
\end{split}
\end{equation}
Taking the mathematical expectation, the inequality \eqref{eq:MAf} becomes
\begin{equation}\label{eq:B1}
    \left| \sum_{\vec{u} \in V_n} W_f(\vec{u}) E(\vec{u}) \right| \leqslant 2^n.
\end{equation}
Note that for the function $\overline{f}(\vec{x}) = 1 \oplus f(\vec{x})$ we have $W_{\overline{f}}(\vec{u}) =
-W_f(\vec{u})$, and due to this the inequality \eqref{eq:B1} for $f \in F_n$ is equivalent to two
inequalities
\begin{equation}\label{eq:B2}
    \sum_{\vec{u} \in V_n} W_f(\vec{u}) E(\vec{u}) \leqslant 2^n
\end{equation}
for $f$ and $\overline{f}$. The $2^{2^n}$ inequalities \eqref{eq:B2} for $f \in F_n$ are referred to as Bell
inequalities. They form a necessary and sufficient condition for $E(\vec{u})$ to be the mathematical
expectations \eqref{eq:E}.

Note that the inequality \eqref{eq:E1} is equivalent to two inequalities
\begin{equation}\label{eq:Bt}
    E(\vec{u}) \leqslant 1 \quad \text{and} \quad -E(\vec{u}) \leqslant 1.
\end{equation}
The first inequality is the Bell inequality \eqref{eq:B2} with the linear function $f(\vec{x}) \equiv
l(\vec{x}) = \langle \vec{u}, \vec{x} \rangle$; the second one corresponds to the affine function
$\overline{l}(\vec{x}) = 1 \oplus \langle \vec{u}, \vec{x} \rangle$, so the trivial inequalities
\eqref{eq:E1} are Bell inequalities with affine functions.

The Bell inequalities \eqref{eq:B2} can be written in the following formal form:
\begin{equation}\label{eq:WfE}
    \langle W_f | E \rangle \leqslant 1,
\end{equation}
where $|E\rangle$ is a (non-normalized) state defined via
\begin{equation}
    |E\rangle = \sum_{\vec{u} \in V_n} E(\vec{u}) |\vec{u}\rangle.
\end{equation}
If $A_i(u_i)$, $i = 1, \ldots, n$ are independent for all $u_i \in \mathbf{Z}_2$, then we have
\begin{equation}
    E(u_1, \ldots, u_n) = q_1(u_1) \ldots q_n(u_n),
\end{equation}
where $q_i(u) = 2\mathbf{P}(A_i(u)=1) - 1$, and $|E\rangle$ is factorizable
\begin{equation}\label{eq:Ef}
    |E\rangle = \bigotimes^n_{i=1} (q_i(0)|0\rangle + q_i(1)|1\rangle).
\end{equation}
Since $|q_i(u)| \leqslant 1$ it is easy to check that any factorizable state \eqref{eq:Ef} satisfy all the
inequalities \eqref{eq:WfE}. But even if there are correlations between $A_i(u_i)$, their mathematical
expectations \eqref{eq:E} satisfy the Bell inequalities \eqref{eq:B2}.

\subsection{Violations of Bell inequalities}

Now let $\hat{A}_i(u_i)$ be Hermitian operators (observables) with the spectra $\{-1, 1\}$ and for a given
n-qubit state $\hat{\varrho}$ let $E(\vec{u})$ be the quantum-mechanical average
\begin{equation}\label{eq:Eq}
    E(\vec{u}) = \langle \hat{A}_1(u_1) \ldots \hat{A}_n(u_n) \rangle_{\hat{\varrho}} =
    \tr \bigl(\hat{\varrho}\hat{A}_1(u_1)\ldots\hat{A}_n(u_n)\bigr).
\end{equation}
If we assume that for the state $\hat{\varrho}$ the result of the measurement of the observable
$\hat{A}_i(u_i)$ can be described by a $\pm 1$-valued random variable $A_i(u_i)$, then the quantum mechanical
average \eqref{eq:Eq} equals to the mathematical expectation \eqref{eq:E} and the quantities $E(\vec{u})$
defined in \eqref{eq:Eq} satisfy the Bell inequalities \eqref{eq:B2}. Such a state $\hat{\varrho}$ is called
classically correlated. Surprisingly, there are quantum states such that the quantities \eqref{eq:Eq} violate
the Bell inequalities (for a definite choice of observables $\hat{A}_i(u_i)$), which means that the
correlations of such states are stronger than any classical ones.

\begin{figure*}
    \subfigure[$n=2$]{\includegraphics[scale=0.9]{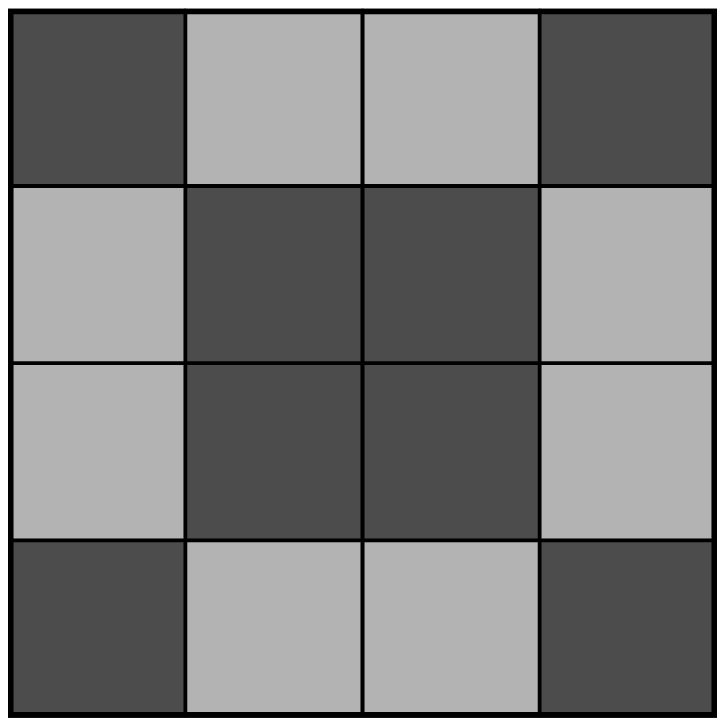}} \hspace*{5mm}
    \subfigure[$n=3$]{\includegraphics[scale=0.9]{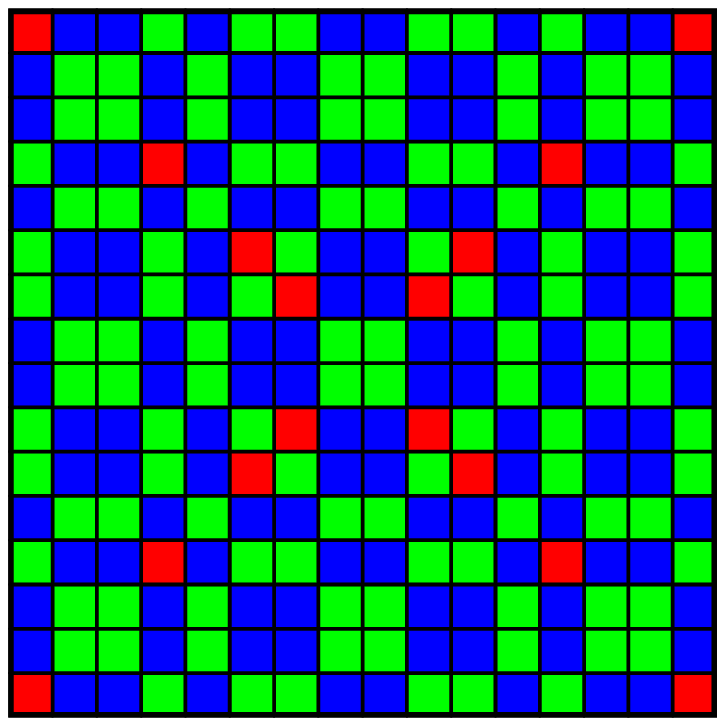}} \\
    \subfigure[$n=4$]{\includegraphics[scale=1.8]{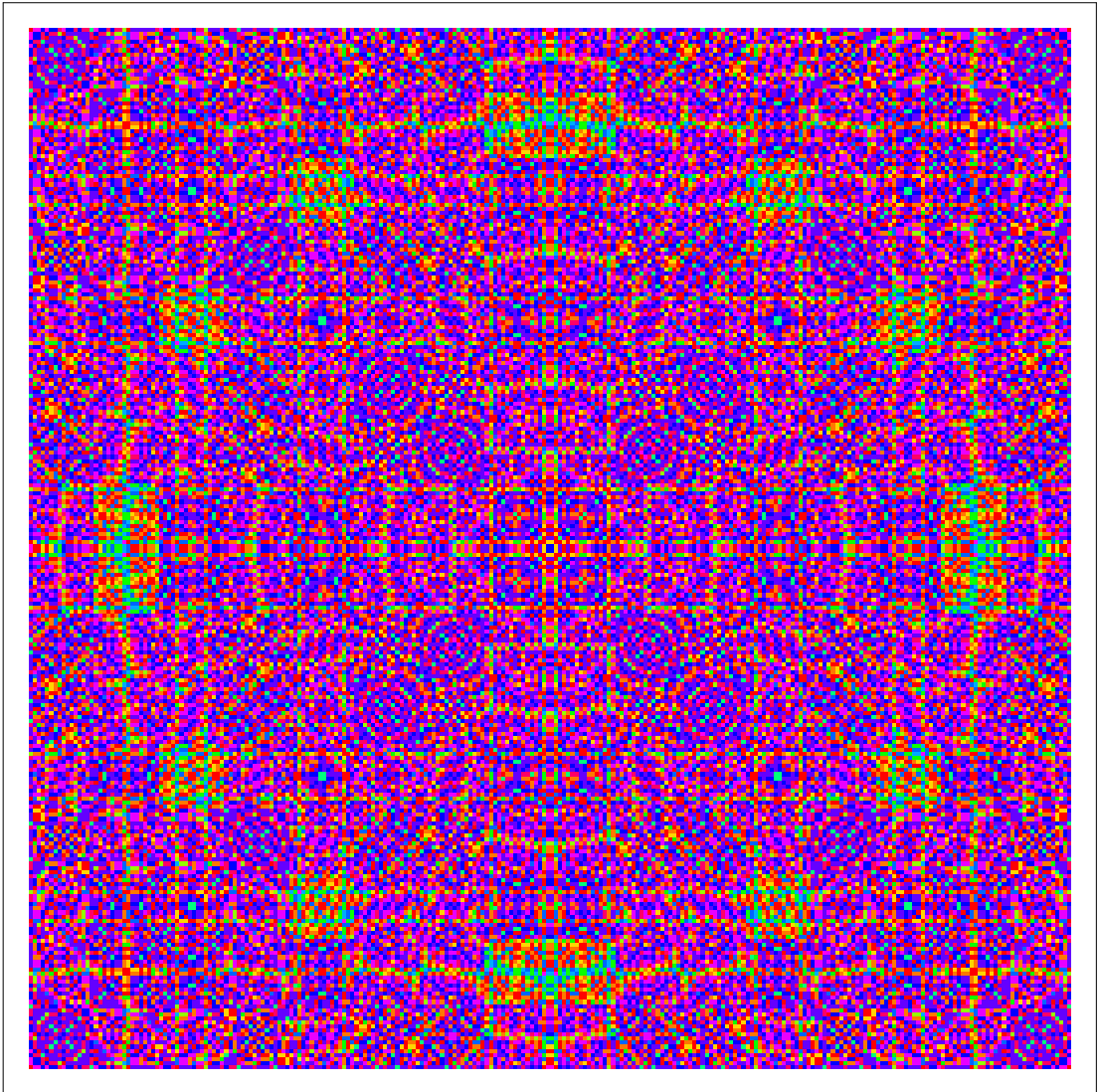}}
\caption{Maximal quantum violation $v_f$ for small $n$.} \label{fig:v}
\end{figure*}

It is interesting to find out up to what extent a given Bell inequality can be violated in quantum case. In
\cite{pra-64-032112} it was shown that the maximal violation $v_f$ of the Bell inequality \eqref{eq:B2}
corresponding to the boolean function $f \in F_n$ reads as
\begin{equation}
    v_f = \frac{1}{2^n} \max_{\varphi_1, \ldots, \varphi_n \in [0, 2\pi]} \left|\sum_{\vec{u} \in V_n}
    e^{i(\vec{\varphi}, \vec{u})} W_f(\vec{u})\right|,
\end{equation}
where $\vec{\varphi} = (\varphi_1, \ldots, \varphi_n)$ and $(\vec{\varphi}, \vec{u}) = \sum^n_{k=1} u_k
\varphi_k$. Using the definition \eqref{eq:WH} of $W_f(\vec{u})$, this expression can be rewritten in the
following form:
\begin{equation}\label{eq:v}
    v_f = \max_{\varphi_1, \ldots, \varphi_n} \left|\sum_{\vec{x} \in V_n} (-1)^{f(\vec{x})}
    \prod^n_{k=1} (-i)^{x_k} t_{x_k}(\varphi_k)\right|,
\end{equation}
where $t_0(\varphi) = \cos\varphi$ and $t_1(\varphi) = \sin\varphi$. In general, it is not an easy
optimization problem and the explicit form of $v_f$ is unknown. The numerically calculated values of $v_f$
for the small $n$ are shown in Fig.~\ref{fig:v}. Studying the numerically obtained results one can conclude
that there is no unique relation between the maximal quantum violation $v_f$ and the nonlinearity $N_f$, but
nevertheless the higher $N_f$ the larger $v_f$ (at least for $n=2, 3, 4$). In other words, the higher
nonlinearity of a boolean function function the stronger maximal quantum violation of the corresponding Bell
inequality.

It was shown that the largest maximal violation reads as
\begin{equation}\label{eq:vm}
    \max_{f \in F_n} v_f = 2^{(n-1)/2}.
\end{equation}
The inequalities on which this upper bound is attained are called Mermin inequalities \cite{prl-65-1838}. In
the next subsection the boolean functions corresponding to these inequalities are explicitly constructed.

\subsection{Mermin inequalities}

Let $A_i(u)$, $i = 1, \ldots, n$, $u \in \mathbf{Z}_2$ be $\pm 1$-valued random variables. For $x, y \in
\mathbf{Z}_2$ define the random variable $M_n(x, y)$ via
\begin{equation}
    M_n(x, y) = \im \prod^n_{k=1}(A_k(x) + i A_k(y)).
\end{equation}
For an odd $n$ the Mermin inequality reads as
\begin{equation}
    \mathbf{M}(M_n(0, 1)) \leqslant 2^{(n-1)/2},
\end{equation}
and for an even $n$ it reads as
\begin{equation}
\begin{split}
    \mathbf{M}\Bigl(&M_{n-1}(0, 1)(A_n(0)+A_n(1)) \\
    + &M_{n-1}(1, 0)(A_n(0)-A_n(1))\Bigr) \leqslant 2^{n/2}.
\end{split}
\end{equation}
After multiplying by a proper number, these inequalities can be written in the form \eqref{eq:B2} with
$W_f(\vec{u}) \equiv M(\vec{u})$ given by
\begin{equation}\label{eq:Wo}
    M(\vec{u}) =
    \begin{cases}
        0 & \text{if}\quad \wt(\vec{u}) \quad \text{is even}, \\
        (-1)^{\frac{\wt(\vec{u})-1}{2}} 2^{\frac{n+1}{2}} & \text{if} \quad \wt(\vec{u}) \quad \text{is odd},
    \end{cases}
\end{equation}
in the case of odd $n$, and by
\begin{equation}\label{eq:We}
    M(\vec{u}) =
    \begin{cases}
        M(\vec{u}')+M(\overline{\vec{u}}') & u_n=0, \\
        M(\vec{u}')-M(\overline{\vec{u}}') & u_n=1,
    \end{cases}
\end{equation}
in the case of even $n$, where $\vec{u}' = (u_1, \ldots, u_{n-1})$ so that $M(\vec{u}')$ is defined by
\eqref{eq:Wo}.

\textit{The function $M(\vec{u})$ defined by \eqref{eq:Wo} or \eqref{eq:We} satisfies the condition
\eqref{eq:WHc}}. Let us first consider the case of odd $n$. For $\vec{v} = \vec{0}$ we have
\begin{equation}
    \sum_{\vec{u} \in V_n} M^2(\vec{u}) = \sum_{\wt(\vec{u})\ \text{is odd}} 2^{n+1} = 2^{n+1} 2^{n-1} = 2^{2n}.
\end{equation}
For $\vec{v} \not= \vec{0}$ first consider the case of odd $\wt(\vec{v})$. Then $\wt(\vec{u} \oplus \vec{v})$
is odd for any $\vec{u}$ with odd $\wt(\vec{u})$ and due to this all terms in the sum in \eqref{eq:WHc} are
zero and the condition \eqref{eq:WHc} is satisfied. The case of even $\wt(\vec{v})$ is more difficult. If
$\wt(\vec{u}) = 2m+1$ then $M(\vec{u}) \sim (-1)^m$ (we omit the common factor $2^{(n+1)/2}$) and
\begin{equation}
    \wt(\vec{u} \oplus \vec{v}) = \wt(\vec{u})+\wt(\vec{v})-2k,
\end{equation}
where $k = \wt(\vec{u} \& \vec{v})$. We have
\begin{equation}
    M(\vec{u} \oplus \vec{v}) \sim (-1)^{m + \wt(\vec{v})/2 -k},
\end{equation}
and the sum in the condition \eqref{eq:WHc} reads as
\begin{equation}\label{eq:Mo}
    \sum_{\vec{u} \in V_n} M(\vec{u}) M(\vec{u} \oplus \vec{v}) \sim \sum_{k}
    \sum_{\wt(\vec{u} \& \vec{v}) = k} (-1)^k.
\end{equation}
The number of terms in the internal sum is
\begin{equation}
    \binom{\wt(\vec{v})}{k} 2^{n-\wt(\vec{v})-1}.
\end{equation}
According to our assumptions $n$ is odd and $\wt(\vec{v})$ is even, hence $n-\wt(\vec{v})-1 \geqslant 0$ and
the sum \eqref{eq:Mo} can be calculated as
\begin{equation}
    \sum_{k} \sum_{\wt(\vec{u} \& \vec{v}) = k} (-1)^k = 2^{n-\wt(\vec{v})-1} (1-1)^{\wt(\vec{v})} = 0.
\end{equation}
This proves that the condition \eqref{eq:WHc} is satisfied also in the case of even $\wt(\vec{v})$. The case
of odd $n$ was completely considered. The proof for the case of even $n$ easily follows from thew definition
\eqref{eq:We}.

Now we will find the boolean functions which correspond to the Mermin inequalities or, in other words, which
maximize $v_f$. First, we will find the boolean function $m \in F_n$ whose Walsch-Hadamard transform is
$M(\vec{u})$. We start with the case of odd $n$. According to \eqref{eq:WHinv} we have
\begin{equation}\label{eq:eq1}
\begin{split}
    &(-1)^{m(\vec{x})} = 2^{-\frac{n-1}{2}} \sum_{\wt(\vec{u})\ \text{is odd}}
    (-1)^{\langle \vec{x}, \vec{u} \rangle + \frac{\wt(\vec{u})-1}{2}} \\
    &= 2^{-\frac{n-1}{2}} \sum_{\substack{1 \leqslant k \leqslant n \\ k \ \text{is odd}}}
    (-1)^{\frac{k-1}{2}} s_k((-1)^{x_1}, \ldots, (-1)^{x_n}),
\end{split}
\end{equation}
where $s_k$ is the $k$-th symmetric polynomial
\begin{equation}
    s_k(z_1, \ldots, z_n) = \sum_{1 \leqslant i_1 < \ldots < i_k \leqslant n} z_{i_1} \ldots z_{i_k}.
\end{equation}
Note that the last sum in \eqref{eq:eq1} can be written as
\begin{equation}
    \sum_{\substack{1 \leqslant k \leqslant n \\ k \ \text{is odd}}}
    (-1)^{\frac{k-1}{2}} s_k(z_1, \ldots, z_n) = (-1)^{\frac{n+1}{2}} \re \prod^n_{l=1} (i - z_l),
\end{equation}
and \eqref{eq:eq1} can be simplified
\begin{equation}
    (-1)^{m(\vec{x})} = -(-2)^{-\frac{n-1}{2}} \re \prod^n_{k=1} (i - (-1)^{x_k}).
\end{equation}
Using the relation
\begin{equation}
    i - (-1)^{x_k} =
    \begin{cases}
        -\sqrt{2} e^{-i \pi/4} & \text{if} \quad x = 0, \\
        \sqrt{2} e^{i \pi/4} & \text{if} \quad x = 1,
    \end{cases}
\end{equation}
we get the equality
\begin{equation}
    (-1)^{m(\vec{x})} = (-1)^{\wt(\vec{x})-\frac{n-1}{2}}
    \re \left((1-i) e^{i\left(\wt(\vec{x}) - \frac{n-1}{2}\right)\frac{\pi}{2}} \right).
\end{equation}
Explicitly the function $m(\vec{x})$ can be written as
\begin{equation}
    m(\vec{x}) =
    \begin{cases}
        0 & \text{if} \quad \wt(\vec{x}) - \frac{n-1}{2} \equiv 0, 3\ (\mathrm{mod}\ 4), \\
        1 & \text{if} \quad \wt(\vec{x}) - \frac{n-1}{2} \equiv 1, 2\ (\mathrm{mod}\ 4).
    \end{cases}
\end{equation}
For any boolean vector $\vec{x} = (x_1, \ldots, x_n) \in V_n$ we have
\begin{equation}
    \bigoplus^n_{k=1} x_k \equiv \wt(\vec{x}), \quad
    \bigoplus_{i < j} x_i x_j \equiv \binom{\wt(\vec{x})}{2} \ (\mathrm{mod} 2),
\end{equation}
from which we get the following result:
\begin{equation}
    m(\vec{x}) =
    \begin{cases}
        \bigoplus\limits^n_{k=1} x_k \oplus \bigoplus\limits_{i < j} x_i x_j &
        \text{if} \ n \equiv 1\ (\mathrm{mod}\ 8), \\
        \bigoplus\limits_{i < j} x_i x_j & \text{if} \ n \equiv 3\ (\mathrm{mod}\ 8), \\
        1 \oplus \bigoplus\limits^n_{k=1} x_k \oplus \bigoplus\limits_{i < j} x_i x_j
        & \text{if} \ n \equiv 5\ (\mathrm{mod}\ 8), \\
        1 \oplus \bigoplus\limits_{i < j} x_i x_j & \text{if} \ n \equiv 7\ (\mathrm{mod}\ 8).
    \end{cases}
\end{equation}
This gives the decomposition of $m$ in the case of odd $n$.

In the case of even $n$ from the definition \eqref{eq:We} we have
\begin{equation}
\begin{split}
    &(-1)^{m(\vec{x})} = 2^{-n} \sum_{\vec{u} \in V_{n-1}} (-1)^{(\vec{x}', \vec{u})} (M(\vec{u}) +
    M(\overline{\vec{u}})) \\
    &+ 2^{-n} \sum_{\vec{u} \in V_{n-1}} (-1)^{(\vec{x}', \vec{u})+x_n} (M(\vec{u}) - M(\overline{\vec{u}})),
\end{split}
\end{equation}
Since $n-1$ is odd, from the previous case the we can conclude that the following relations are valid:
\begin{equation}
\begin{split}
    \sum_{\vec{u} \in V_{n-1}} (-1)^{(\vec{x}', \vec{u})} M(\vec{u}) &= 2^{n-1} (-1)^{m(\vec{x}')}, \\
    \sum_{\vec{u} \in V_{n-1}} (-1)^{(\vec{x}', \vec{u})} M(\overline{\vec{u}}) &=
    2^{n-1} (-1)^{m(\vec{x}') \oplus \wt(\vec{x}) \oplus x_n},
\end{split}
\end{equation}
from which we get
\begin{equation}
\begin{split}
    &(-1)^{m(\vec{x})} = \frac{1}{2} (-1)^{m(\vec{x}')} \\
    &\Bigl(1 + (-1)^{\wt(\vec{x})+x_n} + (-1)^{x_n} - (-1)^{\wt(\vec{x})}\Bigr),
\end{split}
\end{equation}
what is equivalent to the relation
\begin{equation}
    m(\vec{x}) =
    \begin{cases}
        m(\vec{x}') & \text{if} \quad \wt(\vec{x}) \quad \text{is odd}, \\
        m(\vec{x}') \oplus x_n & \text{if} \quad \wt(\vec{x}) \quad \text{is even}.
    \end{cases}
\end{equation}
Explicitly $m(\vec{x})$ reads as
\begin{equation}
    m(\vec{x}) = m(\vec{x}') \oplus \wt(\vec{x}')x_n = m(\vec{x}') \oplus \left(\bigoplus^{n-1}_{k=1}
    x_k\right)x_n.
\end{equation}
This gives the decomposition of $m$ in the case of even $n$.

We see, that in all cases $m(\vec{x})$ is a quadratic form and in all cases its quadratic part is the same
and equals to the symmetric form $s_2$ \eqref{eq:s}. Below the notion for equivalence of Bell inequalities
will be introduced. The maximal quantum violation $v_f$ of equivalent Bell inequalities is the same. It was
shown in \cite{pra-64-032112} that all inequalities that maximize $v_f$ are equivalent. From the
considerations below we can conclude: \textit{a boolean function $f \in F_n$ maximize $v_f$ if and only if it
is of the form}
\begin{equation}\label{eq:mf}
    f(\vec{x}) = a(\vec{x}) \oplus s_2(\vec{x}) = c_0 \oplus \bigoplus^n_{i=1} c_i x_i \oplus
    \bigoplus_{i<j} x_i x_j.
\end{equation}
Affine functions correspond to the trivial Bell inequalities \eqref{eq:Bt} which cannot be violated in
quantum case. Adding $s_2(\vec{x})$ substantially changes their properties: quantum violations become the
largest.

I end this subsection by showing that the boolean functions \eqref{eq:mf}, which correspond to Mermin
inequalities, minimize the uncertainty relation \eqref{eq:ur}, i.e. $U(f)=1$ for all functions of the form
\eqref{eq:mf}. Since the uncertainty is invariant under equivalence of Bell inequalities, it is sufficient to
check the equality $U(f)=1$ only for one boolean function $f$ of the form \eqref{eq:mf}. Let us take $s_2$.
It is easy to see that
\begin{equation}
    NW_{s_2} =
    \begin{cases}
        2^{n-1} & \text{if} \ n \ \text{is odd}, \\
        2^n & \text{if} \ n \ \text{is even}.
    \end{cases}
\end{equation}
Let us calculate $N\Delta_{s_2}$. We have
\begin{equation}
\begin{split}
    \Delta_{s_2}(\vec{u}) &= \sum_{\vec{x} \in V_n} (-1)^{\bigoplus\limits_{i < j} x_i x_j \oplus
    \bigoplus\limits_{i < j} (x_i \oplus u_i)(x_j \oplus u_j)} \\
    &= (-1)^{\bigoplus\limits_{i < j} u_i u_j} \sum_{\vec{x} \in V_n} (-1)^{\langle \tilde{\vec{u}}, \vec{x}
    \rangle},
\end{split}
\end{equation}
where $\tilde{u}_i = \bigoplus_{j \not= i} u_j$. We see that $\Delta_{s_2} \not= 0$ only if $\tilde{\vec{u}}
= 0$, from which it follows that $u_i = \bigoplus^n_{i=1}u_i = \mathrm{const}$. So there are at most two
possibilities: $\vec{u} = (0, \ldots, 0)$ and $\vec{u} = (1, \ldots, 1)$. For an odd $n$ the number $n-1$ is
even and both possibilities give $\Delta_{s_2}(\vec{u})=0$, from which we have $N\Delta_{s_2} = 2$, while for
an even $n$ only the former one, $\vec{u} = (0, \ldots, 0)$, leads to $\Delta_{s_2}(\vec{u})=0$, and in this
case we have $N\Delta_{s_2} = 1$. In both cases we have $U(s_2)=1$.

\section{Classification of Bell inequalities}

The number of Bell inequalities $2^{2^n}$ grows extremely fast with $n$. On the other hand, many of them are
very similar, for example, differ in sign, and due to this such inequalities have similar properties (for
example, they have the same maximal quantum violation). It makes sense to introduce the physically motivated
notion of the equivalence of Bell inequalities and study only the classes of inequivalent inequalities.

Two Bell inequalities (and the two corresponding boolean functions) are said to be equivalent if they can be
obtained from each other by applying the following three kinds of transformations (any number of times, in
any order):
\begin{enumerate}
\renewcommand{\theenumi}{\roman{enumi}}
\renewcommand{\labelenumi}{(\theenumi)}
\item \label{it:p} permuting subsystems,
\item \label{it:o} swapping the outcomes of any observable,
\item \label{it:l} swapping the observables at any site.
\end{enumerate}
Let us study these three kinds of transformations and the equivalence of Bell inequalities in more details.

\subsection{Transformation of the first kind}

The permutation of subsystems, corresponding to a permutation $\pi \in S_n$ ($S_n$ is the symmetric group),
in terms of Bell inequalities reads as: with a Bell inequality \eqref{eq:B2} corresponding to $f \in F_n$ we
associate another inequality with coefficients $W_{f^\prime}(\vec{u}) = W_f(\pi \vec{u})$, where $\pi \vec{u}
\equiv \pi (u_1, \ldots, u_n) = (u_{\pi^{-1}(1)}, \ldots, u_{\pi^{-1}(n)})$. It is easy to see that the
function $W_f(\pi \vec{u})$ satisfies the condition \eqref{eq:WHc} and due to this the function $f'$ exits.
Using the inverse Walsch-Hadamard transform \eqref{eq:WHinv} we can find the relation between $f$ and $f'$:
\begin{equation}
\begin{split}
    (-1&)^{f^\prime(\vec{x})} = \frac{1}{2^n} \sum_{\vec{u} \in V_n} (-1)^{\langle\vec{x}, \vec{u}\rangle} W_f(\pi \vec{u}) \\
    &= \frac{1}{2^n} \sum_{\vec{u} \in V_n} (-1)^{\langle\pi\vec{x}, \pi\vec{u}\rangle} W_f(\pi \vec{u}) \\
    &= \frac{1}{2^n} \sum_{\vec{u} \in V_n} (-1)^{\langle\pi\vec{x}, \vec{u}\rangle} W_f(\vec{u})
    = (-1)^{f(\pi\vec{x})}.
\end{split}
\end{equation}
Here we used the fact that $\langle\pi\vec{x}, \pi\vec{u}\rangle = \langle\vec{x}, \vec{u}\rangle$ for all
$\vec{x}, \vec{y} \in V_n$, $\pi \in \mathcal{S}_n$ and that
\begin{equation}
    \sum_{\vec{u} \in V_n} F(\pi\vec{u}) = \sum_{\vec{u} \in V_n} F(\vec{u}).
\end{equation}
We see that a permutation of subsystems is a permutation of the arguments of the corresponding boolean
function: $f^\prime(\vec{x}) = f(\pi \vec{x})$.

Let us define the map $p_\pi: \mathcal{F}_n \to \mathcal{F}_n$ via
\begin{equation}\label{eq:p}
    (p_\pi f)(\vec{x}) = f(\pi \vec{x}).
\end{equation}
Then any transformation of the type \eqref{it:p} is $p_\pi$ for the appropriate $\pi$. Since $\pi (\sigma
\vec{x}) = (\sigma \pi) \vec{x}$, we have
\begin{equation}
\begin{split}
    (p_\pi p_\sigma f)(\vec{x}) &= (p_\sigma f)(\pi \vec{x}) = f(\sigma (\pi \vec{x})) \\
    &= f((\pi \sigma) \vec{x}) = (p_{\pi \sigma} f)(\vec{x}),
\end{split}
\end{equation}
from which we get the equality
\begin{equation}
    p_\pi p_\sigma = p_{\pi \sigma},
\end{equation}
which is valid for all $\pi, \sigma \in S_n$. In other words, the map $\pi \to p_{\pi}$ is a representation
of $S_n$ in $F_n$.

In terms of the states $|W_f\rangle$ the transformation of this kind is given by the operator $P_\pi$,
defined via:
\begin{equation}
    P_\pi|\vec{x}\rangle = |\pi^{-1} \vec{x}\rangle.
\end{equation}
It is easy to see that $W_{f^\prime}(\vec{u}) = W_f(\pi \vec{u})$ if and only if
\begin{equation}
    |W_{f^\prime}\rangle = P_{\pi} |W_f\rangle.
\end{equation}
This operator is nonlocal --- it permutes subsystems. Note that $P_\pi P_\sigma = P_{\pi\sigma}$.

\subsection{Transformation of the second kind}

Swapping the outcomes of the observable $A_i(0)$ results in the following relations on the coefficients of
Bell inequalities:
\begin{equation}
\begin{split}
    W_{f^\prime}(\ldots, 0, \ldots) &= -W_f(\ldots, 0, \ldots), \\
    W_{f^\prime}(\ldots, 1, \ldots) &= W_f(\ldots, 1, \ldots).
\end{split}
\end{equation}
Similarly, swapping the outcomes of $A_i(1)$ we have
\begin{equation}
\begin{split}
    W_{f^\prime}(\ldots, 0, \ldots) &= W_f(\ldots, 0, \ldots), \\
    W_{f^\prime}(\ldots, 1, \ldots) &= -W_f(\ldots, 1, \ldots).
\end{split}
\end{equation}
These relations can be written as
\begin{equation}
    W_{f^\prime}(\vec{u}) = \pm (-1)^{u_i} W_f(\vec{u}),
\end{equation}
where $+$ corresponds to $A_i(1)$ and $-$ to $A_i(0)$. Swapping the outcomes of several observables $A_i(u)$
with indices $i \in I \subseteq \mathcal{N}_n = \{1, \ldots, n\}$ has the following effect:
\begin{equation}\label{eq:o}
    W_{f^\prime}(\vec{u}) = \pm (-1)^{\langle\vec{e}_I, \vec{u}\rangle} W_f(\vec{u}),
\end{equation}
where all components of $\vec{e}_I \in V_n$ are zero except those with indices in $I$. Any boolean vector
$\vec{y} \in V_n$ can be represented in the form $\vec{e}_I$: for any $\vec{y} \in V_n$ there is $I \subseteq
\mathcal{N}_n$ such that $\vec{y} = \vec{e}_I$. Due to this we characterize transformations of the second
kind by a boolean vector $\vec{y} \in V_n$, and write the relation \eqref{eq:o} as
\begin{equation}\label{eq:o2}
    W_{f^\prime}(\vec{u}) = \pm (-1)^{\langle\vec{y}, \vec{u}\rangle} W_f(\vec{u}).
\end{equation}

From this it is easy to find the relation between $f$ and $f^\prime$ explicitly. For the sign $+$ in
\eqref{eq:o2} we have
\begin{equation}
    (-1)^{f^\prime(\vec{x})} = \frac{1}{2^n} \sum_{\vec{u} \in V_n}
    (-1)^{(\vec{x}\oplus\vec{y}, \vec{u})} W_f(\vec{u}) = (-1)^{f(\vec{x}\oplus\vec{y})},
\end{equation}
so that
\begin{equation}\label{eq:sy}
    f'(\vec{x}) = f(\vec{x}\oplus\vec{y}).
\end{equation}
Analogously, for the sign $-$ in \eqref{eq:o2} we have
\begin{equation}\label{eq:sy2}
    f'(\vec{x}) = f(\vec{x}\oplus\vec{y}) \oplus 1.
\end{equation}

Let us introduce the map $\delta: F_n \to F_n$ and for any $\vec{y} \in V_n$ the map $s_{\vec{y}}: F_n \to
F_n$ via
\begin{equation}\label{eq:deltas}
    (\delta f)(\vec{x}) = f(\vec{x}) \oplus 1, \quad (s_{\vec{y}} f)(\vec{x}) = f(\vec{x} \oplus \vec{y}).
\end{equation}
From the discussion above it follows that any transformation \eqref{it:o} is $s_{\vec{y}}$ or $\delta
s_{\vec{y}}$ for an appropriate $\vec{y} \in V_n$.

Consider the single qubit quantum gate with the matrix
\begin{equation}
    Z =
    \begin{pmatrix}
        1 & 0 \\
        0 & -1
    \end{pmatrix}.
\end{equation}
Due to the equality
\begin{equation}
    \langle \vec{u} | Z^{\vec{y}} |W_f\rangle = (-1)^{\langle \vec{y}, \vec{u} \rangle} W_f(\vec{u})
\end{equation}
the relations \eqref{eq:sy} and \eqref{eq:sy2} are equivalent to the following ones:
\begin{equation}
    |W_{f^\prime}\rangle = \pm Z^{\vec{y}} |W_f\rangle.
\end{equation}
This transformation is local.

\subsection{Transformation of the third kind}

Swapping the observables of the $i$-th site, $A_i(0) \leftrightarrow A_i(1)$, is expressed as
\begin{equation}
\begin{split}
    W_{f^\prime}(\ldots, 0, \ldots) &= W_f(\ldots, 1, \ldots), \\
    W_{f^\prime}(\ldots, 1, \ldots) &= W_f(\ldots, 0, \ldots),
\end{split}
\end{equation}
or in a more compact form as
\begin{equation}
    W_{f^\prime}(\vec{u}) = W_f(\vec{u}\oplus\vec{e}_i).
\end{equation}
Swapping the observables on several sites with indices $J \subseteq \mathcal{N}_n$ reads as
\begin{equation}\label{eq:t}
    W_{f^\prime}(\vec{u}) = W_f(\vec{u} \oplus \vec{e}_J).
\end{equation}
As before, we characterize the transformations under study by a boolean vector $\vec{z} = \vec{e}_J$ and
write the relation \eqref{eq:t} as
\begin{equation}\label{eq:Wt}
    W_{f'}(\vec{u}) = W_f(\vec{u} \oplus \vec{z}).
\end{equation}

It is easy to find the relation between $f$ and $f^\prime$ explicitly:
\begin{equation}
\begin{split}
    &(-1)^{f^\prime(\vec{x})} = \frac{1}{2^n} \sum_{\vec{u} \in V_n} (-1)^{\langle\vec{x}, \vec{u}\rangle} W_f(\vec{u} \oplus \vec{z}) \\
    &= \frac{1}{2^n} \sum_{\vec{u} \in V_n} (-1)^{\langle\vec{x}, \vec{u}\oplus\vec{z}} W_f(\vec{u}) =
    (-1)^{f(\vec{x}) \oplus \langle\vec{x}, \vec{z}\rangle}.
\end{split}
\end{equation}
For any $\vec{z} \in V_n$ let us introduce the map $t_{\vec{z}}: F_n \to F_n$ via
\begin{equation}\label{eq:tz}
    (t_{\vec{z}} f)(\vec{x}) = f(\vec{x}) \oplus \langle\vec{x}, \vec{z}\rangle.
\end{equation}
We see that any transformation \eqref{it:l} is $t_{\vec{z}}$ for an appropriate $\vec{z} \in V_n$.

The relation \eqref{eq:Wt} is equivalent to the following one:
\begin{equation}
    |W_{f^\prime}\rangle = X^\vec{z} |W_f\rangle.
\end{equation}
This is also a local transform.

\subsection{Relations between the transformations}

Let us analyze relations between the maps $p_\pi$, $\delta$, $s_\vec{y}$ and $t_\vec{z}$. It is easy to see
that $\delta$ commutes with all the other maps and satisfies the relation $\delta^2=1$. Composition of two
$s_\vec{y}$ or two $t_\vec{z}$ is again a map of the corresponding kind:
\begin{equation}
    s_{\vec{y}} s_{\vec{y}^\prime} = s_{\vec{y} \oplus \vec{y}^\prime}, \quad
    t_{\vec{z}} t_{\vec{z}^\prime} = t_{\vec{z} \oplus \vec{z}^\prime}.
\end{equation}
We see that the maps $s_\vec{y}$ form a group, isomorphic to the additive group $V_n$; the same is true for
the maps $t_\vec{z}$.

Straightforward calculations show that the following relations are valid for all $\vec{y}, \vec{z} \in V_n$
and $\pi \in S_n$:
\begin{equation}\label{eq:pst}
    p_\pi s_\vec{y} = s_{\pi^{-1}\vec{y}} p_\pi, \quad
    p_\pi t_\vec{z} = t_{\pi^{-1}\vec{z}} p_\pi
\end{equation}
Let us prove only the second relation. We have
\begin{equation}
    (p_\pi t_\vec{z} f)(\vec{x}) = (t_\vec{z} f)(\pi \vec{x}) = f(\pi \vec{x}) \oplus \langle\pi \vec{x}, \vec{z}\rangle.
\end{equation}
On the other hand, we have
\begin{equation}
    (t_{\pi^{-1}\vec{z}} p_\pi f)(\vec{x}) = f(\pi \vec{x}) \oplus \langle\vec{x}, \pi^{-1}\vec{z}\rangle.
\end{equation}
Taking into account that
\begin{equation}
    \langle\pi \vec{x}, \vec{z}\rangle = \langle\pi^{-1} \pi \vec{x}, \pi^{-1} \vec{z}\rangle = \langle\vec{x},
    \pi^{-1}\vec{z}\rangle,
\end{equation}
we see that the last equality in \eqref{eq:pst} is valid.

The maps $s_\vec{y}$ and $t_\vec{z}$ commute modulo the map $\delta$:
\begin{equation}\label{eq:st}
    s_\vec{y} t_\vec{z} =
    \begin{cases}
        t_\vec{z} s_\vec{y} & \text{if}\quad \langle\vec{y}, \vec{z}\rangle = 0, \\
        \delta t_\vec{z} s_\vec{y} & \text{if}\quad \langle\vec{y}, \vec{z}\rangle = 1.
    \end{cases}
\end{equation}
In fact, we have
\begin{equation}
\begin{split}
    (s_\vec{y} t_\vec{z} f)(\vec{x}) &= (t_\vec{z} f)(\vec{x} \oplus \vec{y}) \\
    &= f(\vec{x} \oplus \vec{y}) \oplus \langle\vec{x}, \vec{z}\rangle \oplus
    \langle\vec{y}, \vec{z}\rangle.
\end{split}
\end{equation}
Since
\begin{equation}
    f(\vec{x} \oplus \vec{y}) \oplus \langle\vec{x}, \vec{z}\rangle = (t_\vec{z} s_\vec{y} f)(\vec{x}),
\end{equation}
the equality \eqref{eq:st} is valid.

\subsection{The group $\mathcal{G}_n$}

All the maps $p_\pi$, $\delta$, $s_{\vec{y}}$ and $t_{\vec{z}}$ are invertible and they generate a subgroup
$\mathcal{G}_n$ of the symmetric group $S_{F_n}$ (the group of all one-to-one transformations of $F_n$). From
the relations between $p_\pi$, $\delta$, $s_{\vec{y}}$ and $t_{\vec{z}}$ it follows that any element $\alpha
\in \mathcal{G}_n$ can be represented as
\begin{equation}\label{eq:alpha}
    \alpha = \varepsilon t_\vec{z} p_\pi s_\vec{y}
\end{equation}
in an unique way, where $\varepsilon$ is either $\id: F_n \to F_n$ or $\delta$. In particular,
$|\mathcal{G}_n| = 2^{2n+1} n!$. The element $\alpha$ \eqref{eq:alpha} acts on a function $f \in F_n$ as
\begin{equation}\label{eq:nf}
    (\alpha f)(\vec{x}) =
    \begin{cases}
        f(\pi \vec{x} \oplus \vec{y}) \oplus \langle\vec{x}, \vec{z}\rangle & \text{if} \quad \varepsilon = \id, \\
        f(\pi \vec{x} \oplus \vec{y}) \oplus \langle\vec{x}, \vec{z}\rangle \oplus 1 & \text{if} \quad \varepsilon = \delta.
    \end{cases}
\end{equation}
This action can be written in the following form:
\begin{equation}\label{eq:nf2}
    (\alpha f)(\vec{x}) = f(\pi \vec{x} \oplus \vec{y}) \oplus a(\vec{x}),
\end{equation}
where $a \in A_n$ is an affine function. Now we can formulate the equivalence of boolean functions (and the
corresponding Bell inequalities) as follows: \textit{two boolean functions $f, f' \in F_n$ are equivalent if
there is $\alpha \in \mathcal{G}_n$ such that $f' = \alpha f$.} It is easy to see that it is really an
equivalence on the set $F_n$. Note that $f \sim f'$ if and only if $f \oplus s_2 \sim f' \oplus s_2$ from
which it follows that $S: F_n/\!\!\sim \to F_n/\!\!\sim$, $[f] \mapsto [f \oplus s_2]$ is a nontrivial
involution on the set of the equivalence classes $F_n/\!\!\sim$. The equivalence class $[0]$ is the class of
trivial inequalities and the class $S([0]) = [s_2]$ is the class of the Mermin inequalities.

The group $\mathcal{G}_n$ can be identified with the set
\begin{equation}
    \mathcal{G}_n = \{\id, \delta\} \times V_n \times S_n \times V_n,
\end{equation}
equipped with the product:
\begin{equation}
\begin{split}
    (\varepsilon, &\vec{z}, \pi, \vec{y}) (\varepsilon^\prime, \vec{z}^\prime, \pi^\prime, \vec{y}^\prime) \\
    &= (\varepsilon_0 \varepsilon \varepsilon^\prime, \vec{z} \oplus \pi^{-1}\vec{z}^\prime, \pi \pi^\prime,
    \pi^\prime\vec{y} \oplus \vec{y}^\prime),
\end{split}
\end{equation}
where $\varepsilon_0$ is defined as
\begin{equation}
    \varepsilon_0 =
    \begin{cases}
        \id & \text{if} \quad \langle\vec{y}, \vec{z}^\prime\rangle = 0, \\
        \delta & \text{if} \quad \langle\vec{y}, \vec{z}^\prime\rangle = 1.
    \end{cases}
\end{equation}
The unit element $e$ of this group is
\begin{equation}
    e = (\id, \vec{0}, \id, \vec{0}),
\end{equation}
where the first $\id: F_n \to F_n$ is the identity map on $F_n$ and the second $\id: \mathcal{N}_n \to
\mathcal{N}_n$ is that on $\mathcal{N}_n$. The inverse $(\varepsilon, \vec{z}, \pi, \vec{y})^{-1}$ reads as
\begin{equation}
    (\varepsilon, \vec{z}, \pi, \vec{y})^{-1} = (\tilde{\varepsilon} \varepsilon, \pi\vec{z}, \pi^{-1}, \pi^{-1}\vec{y}),
\end{equation}
where $\tilde{\varepsilon}$ is defined via
\begin{equation}
    \tilde{\varepsilon} =
    \begin{cases}
        \id & \text{if} \quad \langle\vec{y}, \pi\vec{z}\rangle = 0, \\
        \delta & \text{if} \quad \langle\vec{y}, \pi\vec{z}\rangle = 1.
    \end{cases}
\end{equation}

The elements $(\varepsilon, \vec{z}, \id, \vec{0})$ form a normal subgroup $\mathcal{U}^{(1)}_n \lhd
\mathcal{G}_n$. The elements $(\id, \vec{0}, \pi, \vec{y})$ form a (not normal) subgroup $\mathcal{J}_n$,
usually referred to as Jevons group. The group $\mathcal{G}_n$ is the semidirect product of
$\mathcal{U}^{(1)}_n$ and $\mathcal{J}_n$:
\begin{equation}
    \mathcal{G}_n = \mathcal{U}^{(1)}_n \leftthreetimes \mathcal{J}_n.
\end{equation}
The elements $(\id, \vec{0}, \id, \vec{y})$ form a normal subgroup of $\mathcal{J}_n$, which is isomorphic to
product of $n$ copies of the cyclic group $C_2$ of second order. The elements $(\id, \vec{0}, \pi, \vec{0})$
form a (not normal) subgroup of $\mathcal{J}_n$ which is isomorphic to the permutation group $S_n$. The
Jevons group $\mathcal{J}_n$ is the semidirect product of $C^n_2$ and $S_n$:
\begin{equation}
    \mathcal{J}_n = C^n_2 \leftthreetimes S_n,
\end{equation}
so we have the following decomposition of $\mathcal{G}_n$:
\begin{equation}
    \mathcal{G}_n = \mathcal{U}^{(1)}_n \leftthreetimes (C^n_2 \leftthreetimes S_n).
\end{equation}

In terms of vectors $|W_f\rangle$ the equivalence can be expressed as: two boolean functions $f, f^\prime \in
\mathcal{F}_n$ are equivalent if and only if there are $\pi \in S_n$ and $\vec{y}, \vec{z} \in V_n$ such that
\begin{equation}
    |W_{f^\prime}\rangle = \pm P_\pi Z^\vec{y} X^\vec{z} |W_f\rangle.
\end{equation}
This gives another interpretation of the action of the group $\mathcal{G}_n$ on the set $F_n$.

\begin{figure*}
    \subfigure[$n=2$]{\includegraphics[scale=0.9]{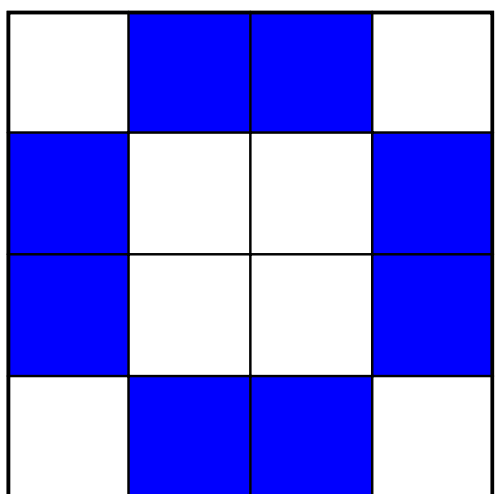}} \hspace*{5mm}
    \subfigure[$n=3$]{\includegraphics[scale=0.9]{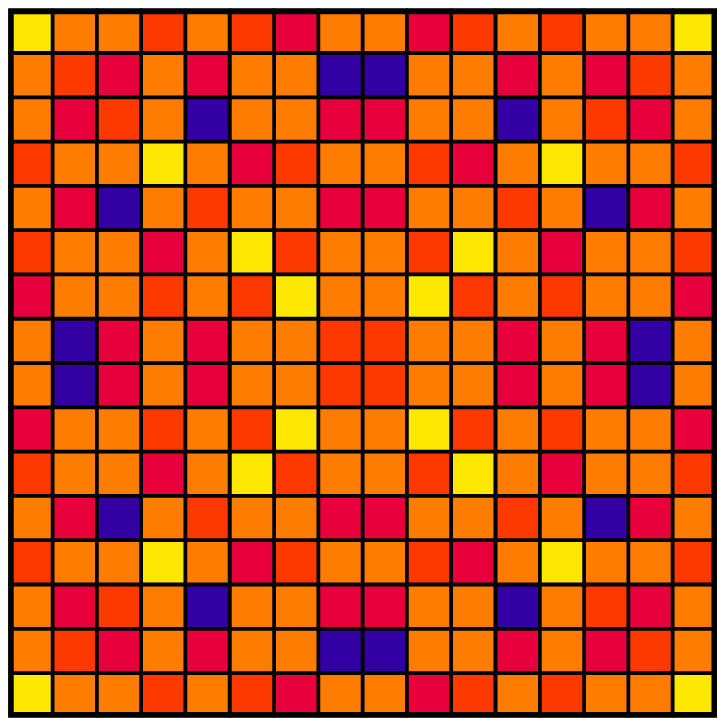} \label{sfig:bell-b}} \\
    \subfigure[$n=4$]{\includegraphics[scale=1.8]{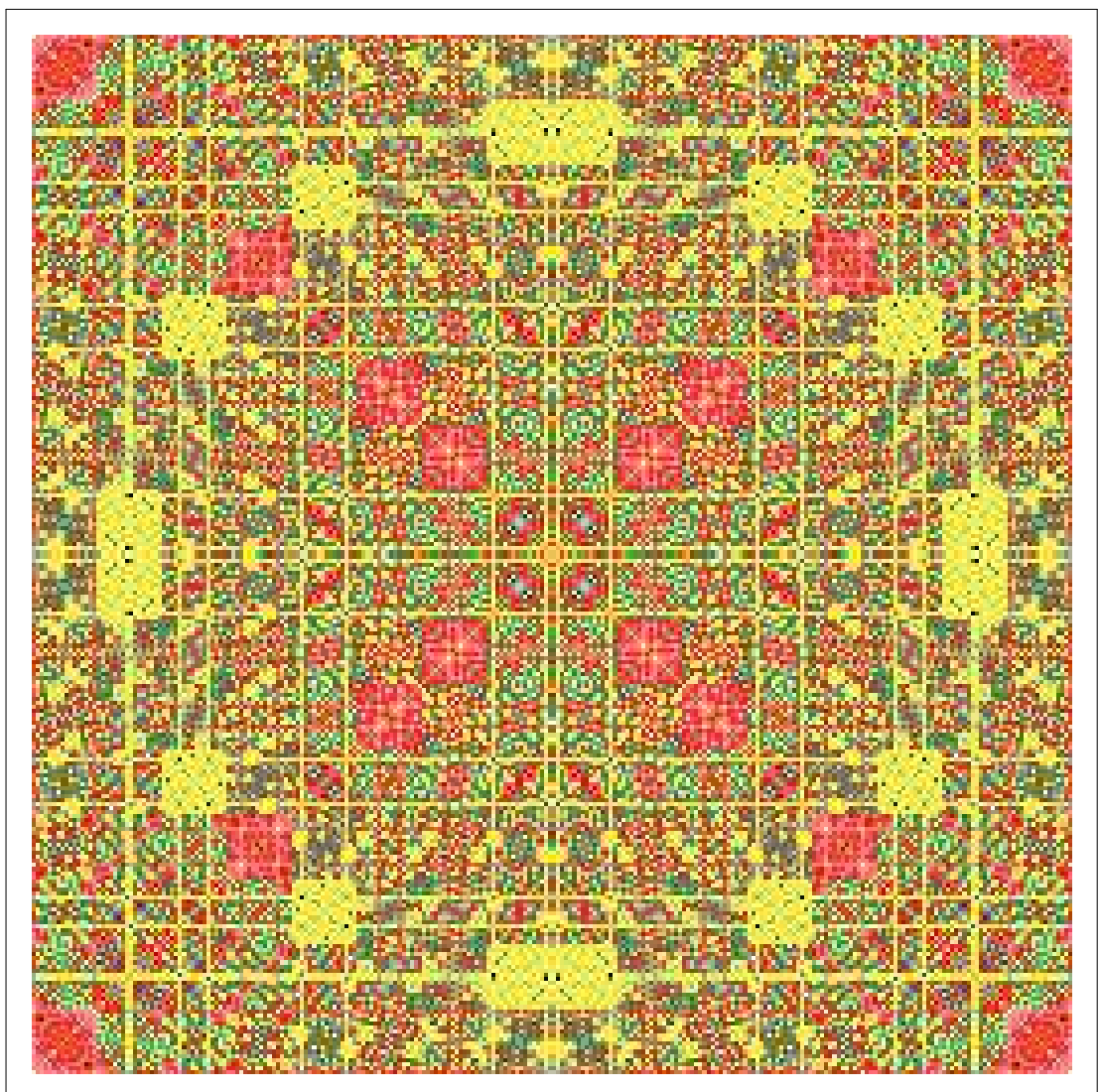} \label{sfig:bell-b2}}
\caption{The classes of equivalence of Bell inequalities for small number of qubits.}\label{fig:bell}
\end{figure*}

\begin{figure*}
    \subfigure{\includegraphics{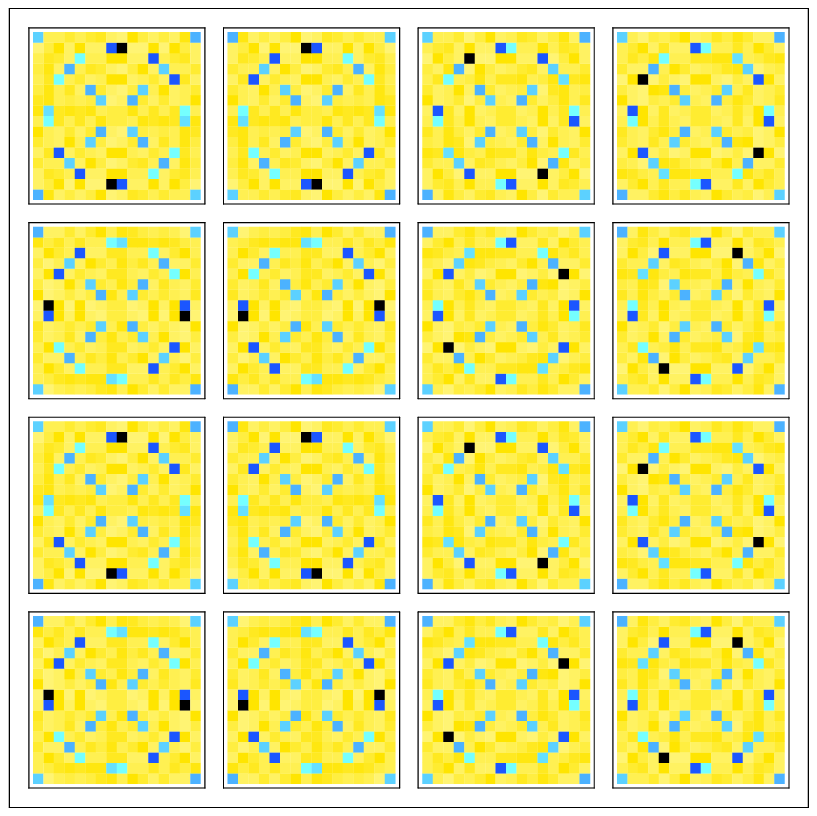}} \hspace*{5mm}
    \subfigure{\includegraphics{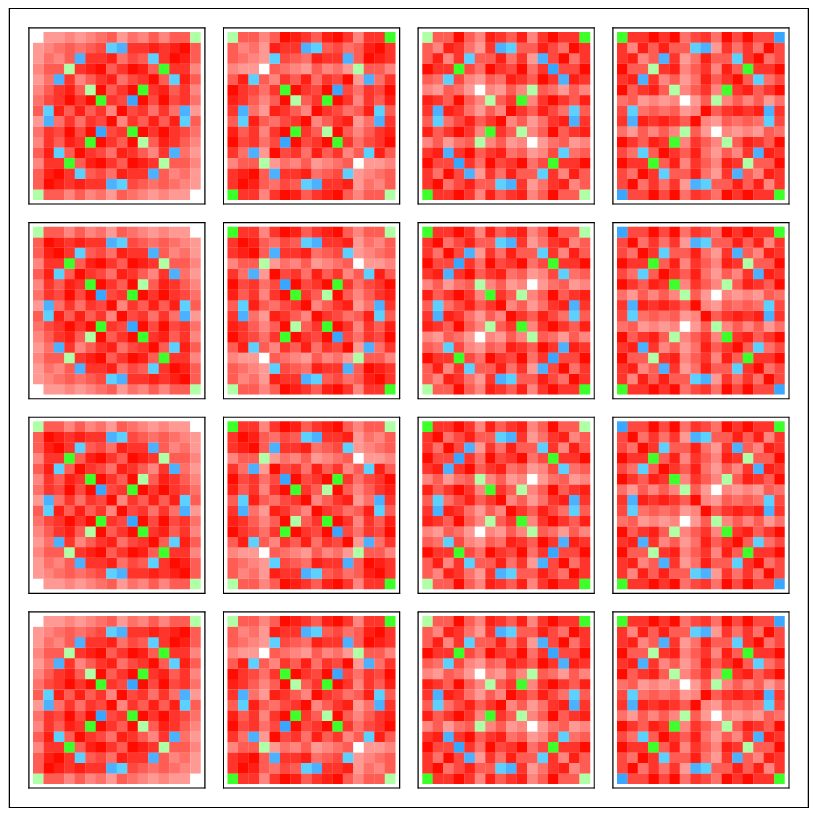}}
\caption{The subsquares.} \label{fig:sq}
\end{figure*}

\begin{wrapfigure}[10]{r}[34pt]{48mm}
    \vspace*{-3mm}
    \hspace*{-3mm} \includegraphics[scale=0.5]{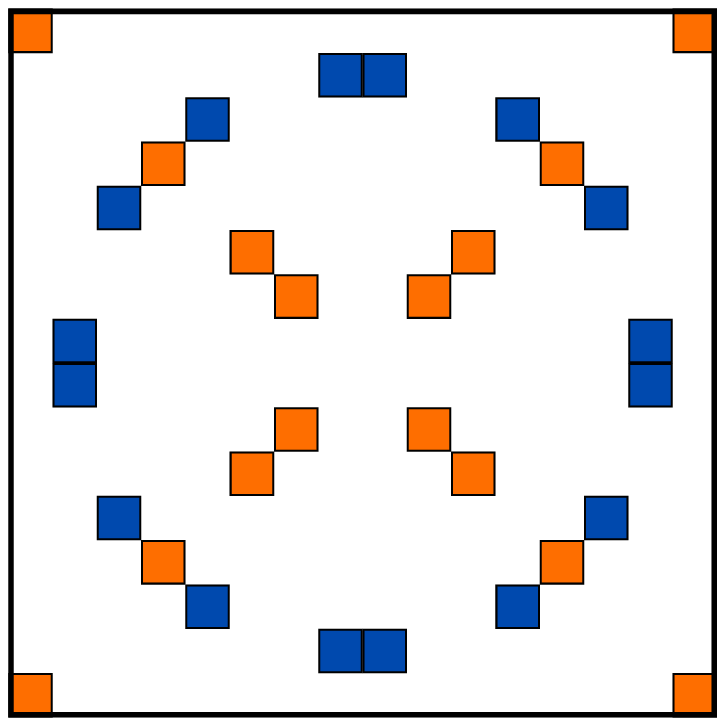}
\end{wrapfigure}

The classes of equivalence of Bell inequalities for small number of qubits are shown in Fig.~\ref{fig:bell}.
For $n=2$ there are $N_2 = 2$ classes of equivalence, $N_3 = 5$ and $N_4 = 39$. The small squares, outlined
in this figure, are shown in Fig.~\ref{fig:sq}. They have the structure similar with that of the
Fig.~\ref{sfig:bell-b}. The blue subsquares (which form a circle) correspond to Mermin inequalities, and the
orange ones (which form a cross) correspond to trivial inequalities. The same picture is repeated in these
subsquares, showing a kind of fractal behavior. The \textsl{Mathematica} code with which these figures were
obtained is presented in the Appendix A.

The figures \ref{fig:bell}, as well as the figures \ref{fig:u} and \ref{fig:v} were obtained numerically. The
analytical expressions for the number of equivalence classes and for the maximal quantum violation $v_f$ of
the equivalence class defined by a boolean function $f \in F_n$ are unknown. In the next two subsections a
partial approach to this problem will be presented.

\subsection{P\'{o}lya theory}

Consider two finite sets $X$ and $Y$. The notation $Y^X$ is used for the set of all the functions $f: X \to
Y$. The set $Y^X$ is finite and has $|Y|^{|X|}$ elements, which motivates the notation. Let $G$ be a finite
group and let $\alpha: G \to S_X$ be a homomorphism of $G$ to the symmetric group on $X$, i.e. an action of
$G$ on $X$. This means that for any $g \in G$ the map
\begin{equation}
    \alpha_g = \alpha(g): X \to X
\end{equation}
is a permutation of $X$ and the relation
\begin{equation}
    \alpha_g \alpha_{g^\prime} = \alpha_{g g^\prime}
\end{equation}
is valid for all $g, g^\prime \in G$.

Two functions $f, f^\prime: X \to Y$ are said to be equivalent with respect to the action $\alpha$, $f
\sim_\alpha f^\prime$, if there is $g \in G$ such that $f' = f \circ \alpha_g$, i.e. if the relation
\begin{equation}\label{eq:equiv}
    f^\prime(x) = f(\alpha_g x)
\end{equation}
is valid for all $x \in X$. What is the number $N$ of equivalence classes? For example, if the action
$\alpha$ is trivial, i.e. if $\alpha_g = \id_X$ for all $g \in G$, then $f \sim_\alpha f^\prime$ if and only
if $f = f'$, which means that each equivalence class consists of a single element and there are $|Y|^{|X|}$
equivalence classes.

To calculate the number of equivalence classes we need the notion of the cycle index of the group $G$. Each
element $g \in G$ defines an equivalence $\sim_g$ on the set $X$ via
\begin{equation}
    x \sim x^\prime \quad \text{if} \quad x^\prime = \alpha_g(x).
\end{equation}
Each equivalence class has the form of a cycle
\begin{equation}
    z_g(x) = \{x, \alpha_g(x), \ldots, \alpha^{k(x)-1}_g(x)\}
\end{equation}
of some length $k(x)$ (so that $\alpha^{k(x)}_g(x) = x$). Different cycles $z_g(x)$ do not intersect. For any
fixed $g \in G$ the set $X$ can be represented as a disjoint union of cycles
\begin{equation}
    X = \bigcup_i z_g(x_i), \quad z_g(x_i) \cap z_g(x_j) = \varnothing \quad \text{if} \quad i
    \not= j.
\end{equation}
Let $c_k(g)$ be the number of the cycles of the length $k$ in this decomposition, $k = 1, \ldots, n = |X|$.
It is clear that the numbers $c_1(g), \ldots, c_n(g)$ satisfy the following relation
\begin{equation}
    \sum^n_{k=1} k c_k(g) = n.
\end{equation}
The cycle index of $G$ with respect to the action $\alpha$ is the polynomial $Z_G(x_1, \ldots, x_n)$ of $n$
variables defined as
\begin{equation}
    Z_G(x_1, \ldots, x_n) = \frac{1}{|G|} \sum_{g \in G} x^{c_1(g)}_1 \ldots x^{c_n(g)}_n.
\end{equation}
A very special case of the P\'{o}lya theorem is the following statement: \textit{for the number of the
equivalence classes (with respect to the equivalence \eqref{eq:equiv}) we have}
\begin{equation}\label{eq:nce}
    N = Z_G(|Y|, \ldots, |Y|).
\end{equation}

Now consider a more general situation. where there is an action not only on $X$ but also an action $\beta: H
\to S_Y$ of the group $H$ on $Y$. The equivalence of functions given by \eqref{eq:equiv} can be extended as
follows: $f, f^\prime: X \to Y$ are said to be equivalent, $f \sim f^\prime$, if there are $g \in G$ and $h
\in H$ such that the diagram
\begin{equation}
\begin{CD}
    X @>\alpha_g>> X \\
    @VVf'V       @VVfV \\
    Y @>\beta_h>> Y
\end{CD}
\end{equation}
is commutative, i.e. if the relation
\begin{equation}\label{eq:equiv2}
    \beta_h f^\prime(x) = f(\alpha_g x)
\end{equation}
is valid for all $x \in X$. In this case the number $N$ of equivalence classes can be expressed in terms of
the cycle indices $Z_G(x_1, \ldots, x_n)$ and $Z_H(y_1, \ldots, y_m)$ as follows:
\begin{equation}
    N = Z_G\left(\frac{\partial}{\partial z_1}, \ldots, \frac{\partial}{\partial z_n}\right)
    Z_H(e^{s_1(\vec{z})}, \ldots, e^{m s_1(\vec{z})}),
\end{equation}
where $s_1(\vec{z}) = \sum^n_{j=1} z_j$ and the derivatives are taken at $z_1 = \ldots = z_n = 0$. This
expression can be transformed to another form, sometimes more suitable for calculations:
\begin{equation}\label{eq:nce2}
    N = \frac{1}{|H|} \sum_{h \in H} Z_G(c_1(h), c_1(h)+2c_2(h), \ldots),
\end{equation}
where the $i$-th argument is equal to
\begin{equation}
    \sum_{j | i} j c_j(h).
\end{equation}
It easy to see that the previous case (where there is a action only on $X$) is a special case of this more
general situation when the action of $H$ on $Y$ is trivial. In fact, in such a case we have
\begin{equation}
    c_1(h) = |Y|, \quad c_k(h) = 0 \quad \text{if} \quad k>1,
\end{equation}
for all $h \in H$, and from \eqref{eq:nce2} we get the relation \eqref{eq:nce}:
\begin{equation}
    N = \frac{1}{|H|} \sum_{h \in H} Z(|Y|, \ldots, |Y|) = Z(|Y|, \ldots, |Y|).
\end{equation}

If the equivalence on the set $Y^X$ can be represented in the form \eqref{eq:equiv2} then the number of
equivalence classes can be calculated according to \eqref{eq:nce2} (or \eqref{eq:nce}). Unfortunately, not
any equivalence on $Y^X$ can be represented in the form \eqref{eq:equiv2}. In such cases one must find other
ways to solve the problem.

\subsection{Classification with respect Jevon's group $\mathcal{J}_n$}

In our case $X = V_n$ and $Y = \mathbf{Z}_2$. The equivalence \eqref{eq:nf2} cannot be represented directly
in the form \eqref{eq:equiv2}, so let us start with a simpler case. Consider the action of $\mathcal{J}_n$ on
$V_n$, which is the reduction of the action of $\mathcal{G}_n$ to its subgroup $\mathcal{J}_n$. The cycle
index of the group $\mathcal{J}_n$ was calculated in \cite{umtn-04879-3-T} and it is given by the following
complicated expression:
\begin{widetext}
\begin{equation}\label{eq:Z}
    Z_{\mathcal{J}_n}(x_1, \ldots, x_{2^n}) = \sum_{\vec{c}}
    \left(\frac{1}{\prod\limits^n_{i=1} c_i! (2i)^{c_i}} \bigotimes^n_{i=1}
    \left(\prod_{d|i} x^{a(d)}_d + \prod_{d|2i, d \nmid i} x^{b(d)}_d\right)^{\otimes c_i}\right),
\end{equation}
\end{widetext}
where the sum is over all vectors $\vec{c} = (c_1, \ldots, c_n)$ with nonnegative integer components such
that
\begin{equation}
    \sum^n_{k=1} k c_k = n.
\end{equation}
The function $a(k)$ is defined for any integer $k \geqslant 1$ via
\begin{equation}\label{eq:a}
    a(k) = \frac{1}{k} \sum_{d | k} 2^d \mu\left(\frac{k}{d}\right),
\end{equation}
and $b(2k)$ is defined for positive even integer argument $2k$, $k \geqslant 1$, via
\begin{equation}\label{eq:b}
    b(2k) = \frac{1}{2k} \sum_{d | 2k, d \nmid k} 2^{d/2} \mu\left(\frac{2k}{d}\right),
\end{equation}
with $\mu(m)$ being the M\"{o}bius function
\begin{equation}
    \mu(m) =
    \begin{cases}
        1 & \text{if} \quad m=1, \\
        (-1)^k & \text{if} \quad m = p_1 \ldots p_k, \\
        0 & \text{in other cases},
    \end{cases}
\end{equation}
where $p_1, \ldots, p_k$ are different prime numbers.

\begin{table}
\begin{tabular}{|c|c|c|c|} \hline
$n$ & $Z_{\mathcal{J}_n}(x_1, \ldots, x_{2^n})$ & $\overline{N}_n$ & $N_n$ \\
\hline 1 & \parbox{50mm}{\begin{displaymath}\frac{1}{2}(x^2_1 + x_2)\end{displaymath}} & $2$ & $1$ \\
\hline 2 & \parbox{50mm}{\begin{displaymath}\frac{1}{8}(x^4_1 + 3 x^2_2 + 2 x^2_1 x_2 + 2
x_4)\end{displaymath}} & $4$ & $2$ \\
\hline 3 & \parbox{50mm}{\begin{displaymath}\begin{split}\frac{1}{48}(x^8_1 &+ 13 x^4_2 + 8 x^2_1
x^2_3 + 8 x_2 x_6 \\ &+ 6 x^4_1 x^2_2 + 12 x^2_4)\end{split}\end{displaymath}} & $14$ & $5$ \\
\hline 4 & \parbox{50mm}{\begin{displaymath}\begin{split}&\frac{1}{384}(x^{16}_1 + 51 x^8_2 + 48 x^2_1 x_2
x^3_4 \\ &+ 48 x^2_8 + 12 x^8_1 x^4_2 + 84 x^4_4 \\ &+ 12 x^4_1 x^6_2 + 32 x^4_1 x^4_3 +
96 x^2_2 x^2_6)\end{split}\end{displaymath}} & $222$ & $39$ \\
\hline 5 & \parbox{50mm}{\begin{displaymath}\begin{split}&\frac{1}{3840}(x^{32}_1 + 231 x^{16}_2 + 20
x^{16}_1 x^8_2 \\ &+ 520 x^8_4 + 80 x^8_1 x^8_3 + 720 x^4_2 x^4_6 \\ &+ 160 x^4_1 x^2_2 x^4_3 x^2_6 + 320
x^2_4 x^2_{12} \\ &+ 240 x^4_1 x^2_2 x^6_4 + 480 x^4_8 + 240 x^4_2 x^6_4 \\ &+ 60 x^8_1
x^{12}_2 + 384 x^2_1 x^6_5 + 384 x_2 x^3_{10})\end{split}\end{displaymath}} & $616126$ & $22442$ \\
\hline
\end{tabular}
\caption{Cycle index $Z_{\mathcal{G}_n}$ for small $n$.}\label{tbl:1}
\end{table}

The cross-product is defined as follows. For powers of variables we have
\begin{equation}\label{eq:x}
    x^n_p \otimes x^m_q = x^{n m \gcd(p, q)}_{\lcm(p, q)},
\end{equation}
where $\gcd(p, q)$ and $\lcm(p, q)$ are the greatest common divisor and the least common multiple of integers
$p$ and $q$ respectively. For two monomials the cross-product is defined via
\begin{equation}
    (x^{n_1}_{p_1} \ldots x^{n_k}_{p_k}) \otimes (x^{m_1}_{q_1} \ldots x^{m_l}_{q_l}) =
    \prod^k_{i=1} \prod^l_{j=1} (x^{n_i}_{p_i} \otimes x^{m_j}_{q_j}),
\end{equation}
and then extended for arbitrary polynomials by bilinearity. For example, let us calculate the cross-product
$(x^2_1 x^3_2) \otimes (x^4_3 x^5_4)$:
\begin{equation}
\begin{split}
    &(x^2_1 x^3_2) \otimes (x^4_3 x^5_4) = (x^2_1 \otimes x^4_3) (x^2_1 \otimes x^5_4) (x^3_2 \otimes x^4_3) (x^3_2 \otimes x^5_4) \\
    &= x^{2 \cdot 4 \cdot 1}_3 x^{2 \cdot 5 \cdot 1}_4 x^{3 \cdot 4 \cdot 1}_6 x^{3 \cdot 5 \cdot 2}_4 = x^8_3 x^{40}_4 x^{12}_6.
\end{split}
\end{equation}
The cycle index $Z_{\mathcal{J}_n}$ for small $n$ is shown in the second column of the table \ref{tbl:1}.

Now let us add an action on the set $Y = \mathbf{Z}_2$. Let $H = C_2 = \{1, \tau\}$ be the cyclic group of
the second order. The action $\beta$ we define as: $\beta_1 = \id_{\mathbf{Z}_2}$ and $\beta_\tau$ being the
logical \textsl{NOT}, $\beta_\tau(0) = 1$ and $\beta_\tau(1)=0$. For cycle lengthes we have
\begin{equation}
    c_1(1)=2, \quad c_2(1)=0, \quad c_1(\tau)=0, \quad c_2(\tau)=1.
\end{equation}
According to \eqref{eq:nce2} for the number of equivalence classes we have the following expression (see also
\cite{umtn-04879-4-T}):
\begin{equation}
    \overline{N}_n = \frac{1}{2}\Bigl(Z_G(2, \ldots, 2) + Z_G(0, 2, 0, 2, \ldots)\Bigr).
\end{equation}
These numbers are shown in the third column of the table \ref{tbl:1}. The last column of the table shows the
number $N_n$ of equivalence classes with respect to the equivalence under study. The number $N_5$ was taken
from \textsl{www.ii.uib.no/~larsed/boolean/}. The numbers $N_n$ for $n > 5$ are unknown.

\section{Conclusion}

In conclusion, the relation between the boolean functions theory and the general Bell inequalities for
$n$-qubits is established. The classification of Bell inequalities with respect to the Jevons group is
obtained, which is a weaker result then the problem posed in \cite{pra-64-032112}. Nevertheless, to my
knowledge it is the only approach to the more general classification. This approach is based on the works
\cite{umtn-04879-3-T, umtn-04879-4-T} done for computer logic circuits theory, which shows the connection
between quite different problems
--- qubit system description and computer logic circuit design.

There are still many unsolved problems. Two the most important ones are:
\begin{enumerate}
\renewcommand{\theenumi}{\roman{enumi}}
\renewcommand{\labelenumi}{(\theenumi)}

\item classification of Bell inequalities (or boolean functions) with respect to the group $\mathcal{G}_n$,

\item characterization of the maximal quantum violation $v_f$ of a given boolean function $f$ in terms of
properties of $f$, in particular, finding the relation between $v_f$ and the nonlinearity $N_f$.
\end{enumerate}
And, of course, a very interesting question --- how the ideas from different applications of the boolean
functions theory (not only to cryptography or computer logic circuits design) can be used in quantum
information theory.

\appendix

\section{The code}
In this Appendix I present the \Mathematica code Figs. \ref{fig:o}, \ref{fig:u}, \ref{fig:v}, \ref{fig:bell}
and Table \ref{tbl:1} were obtained with. First of all, the set $V_n$ can be coded as
\begin{equation}\label{eq:AV}
    \In \mathtt{V[n\_] := Tuples[\{0, 1\}, n]}
\end{equation}
The \Mathematica function $\mathrm{IntegerDigits[n, b, l]}$ gives the list of the base-$b$ digits of $n$,
padding it on the left if necessary to give a list of length $l$. Using this function, one can code $V_n$ in
another way as
\begin{equation}
    \mathtt{V[n\_] := IntegerDigits[\#, 2, n]\& /@ Range[0, 2^n-1]} \nonumber
\end{equation}
This code works a few times more slowly, but this approach can be useful if it necessary to construct only
some part of $V_n$, not the whole set $V_n$. We know that $F_n \simeq V_{2^n}$, but since $2^{2^6} >
10^{19}$, the simple definition $\mathtt{F[n\_] :=\ V[2^n]}$ will not work for $n \geqslant 6$.

A general boolean function \eqref{eq:fcoeff} can be coded as follows:
\begin{widetext}
\begin{equation}
    \In \mathtt{f[c\_, S\_] :=
    Function[x, Mod[c.(Apply[Times, \#]\& /@\ (Part[x, \#]\& /@\ S)), 2]]}
\end{equation}
\end{widetext}
\noindent where $\mathtt{J}$ is the list of all the multi-indices $(i_1, \ldots, i_k)$ for which the
coefficients $c_{i_1, \ldots, i_k} \not= 0$ and $\mathtt{c}$ the list of these coefficients. Note that the
$\mathtt{Length}$-es of $\mathtt{c}$ and $\mathtt{J}$ must be the same. For example, the expression
\begin{equation}\label{eq:Ac}
    \mathtt{f[c, Subsets[Range[n], \{m\}]]}
\end{equation}
gives a homogeneous polynomial of the degree $m$ (except the case when all elements of $\mathtt{c}$ are zero;
when all elements of the list $\mathtt{c}$ are $1$ then \eqref{eq:Ac} is the $m$-th symmetric polynomial
\eqref{eq:s}:
\begin{widetext}
\begin{equation}
    \mathtt{f[Table[1, \{Binomial[n, m]\}], Subsets[Range[n], \{m\}]][\{x_1, \ldots, x_n\}] =
    s_m(x_1, \ldots, x_n)}
\end{equation}
\end{widetext}
\noindent Similarly, the expression
\begin{equation}
    \mathtt{f[c, Subsets[Range[n], m]]}
\end{equation}
is a polynomial of the degree not greater then $m$.

The Walsch-Hadamard transform can be coded as
\begin{widetext}
\begin{equation}
    \In \mathtt{WHT :=\ Function[u, Plus\ @@\ (Function[x, (-1)^{\#[x]+(x.u)}] /@\
    V[Length[u]])]\&}
\end{equation}
\end{widetext}
\noindent This method of calculation the Walsch-Hadamard transform is very simple and quite ineffective. In
Appendix B a much better method is presented. The only disadvantage of that method is the fact that it works
much faster only when it is necessary to calculate all the numbers $W_f(\vec{u})$, $\vec{u} \in V_n$
simultaneously, and it cannot be applied to calculate only one number $W_f(\vec{u})$ for a given $\vec{u} \in
V_n$. $\mathtt{WHT}$ is a functional: $\mathtt{WHT[f]}$ is a function and $\mathtt{WHT[f][\{u_1, \ldots,
u_n\}]}$ is its value $W_f(\vec{u})$ at $\vec{u} = (u_1, \ldots, u_n)$. The inverse Walsch-Hadamard transform
can be coded as
\begin{widetext}
\begin{equation}
    \In \mathtt{WHIT :=\ Function[x, (1 - 2^{-Length[x]}
    Plus\ @@\ (Function[u, (-1)^{x.u}\ \#[u]] /@\ V[Length[x]]))/2] \&}
\end{equation}
\end{widetext}
\noindent Like $\mathtt{WHT}$, it is also a functional.

The autocorrelation \eqref{eq:Delta} of two boolean functions can be coded as
\begin{widetext}
\begin{equation}
    \In \mathtt{\Delta :=\ Function[u, Plus\ @@\
    (Function[x, (-1)^{\#1[x] + \#2[Mod[\#, 2]\& /@\ (x+u)]}] /@\ V[Length[u]])] \&}
\end{equation}
\end{widetext}
\noindent It is a functional of two arguments: the expression $\mathtt{\Delta[f, g][\{u_1, \ldots, u_n\}]}$
is the value $\Delta_{f, g}(\vec{u})$ at $\vec{u} = (u_1, \ldots, u_n)$.

The uncertainty \eqref{eq:U} can be coded as
\begin{widetext}
\begin{equation}
    \In \mathtt{U[n\_] :=\ Function[f, 2^{-n} Count[WHT[f] /@\ V[n], \_?(\# != 0 \&)]
    Count[\Delta[f, f] /@\ V[n], \_?(\# != 0 \&)]]}
\end{equation}
\end{widetext}
\noindent For a function $f \in F_n$ the expression $\mathtt{U[n][f]}$ gives the uncertainty $U(f)$ of $f$.
To calculate $U(f)$ for all $f \in F_n$ we need a way to define $f$ given its number $B_n(f)$, $0 \leqslant
B_n(f) <2^{2^n}$. The following code solves this problem:
\begin{widetext}
\begin{equation}
    \In \mathtt{itof[n\_, B\_] :=\ Part[IntegerDigits[B,2, 2^n], 2^n - FromDigits[\#, 2]] \&}
\end{equation}
\end{widetext}
\noindent Given $0 \leqslant B < 2^{2^n}$, the expression $\mathtt{itof[n, B]}$ is the corresponding boolean
function, to which one can apply the syntaxis $\mathtt{itof[n, B][\{x_1, \ldots, x_n\}]}$. We can visualize
boolean functions with respect to their uncertainty using the code
\begin{widetext}
\begin{equation}
    \In \mathtt{ut[n\_] :=\ Partition[Table[U[n][itof[n, B]], \{B, 0, 2^{2^n}-1\}], 2^{2^{n-1}}]}
\end{equation}
\end{widetext}
\noindent In \Mathematica this table can be immediately plotted with the $\mathtt{ArrayPlot}$ function, what
was doe for the case of $n=4$, for the other two cases I used a simple script to generate \textsl{pstricks}
code from this table and then compiled it with \LaTeX (\textsl{pstricks} code produced huge pictures in the
case of $n=4$). In this way Fig.~\ref{fig:u} was obtained. The code
\begin{equation}
    \mathtt{ut[n] //Flatten //Sort //Split //Length}
\end{equation}
\noindent gives the number of different values of the uncertainty. For $n=2$ it is $1$ (all $16$ boolean
functions have the same uncertainty $1$), for $n=3$ it is $2$ (the values are $1$ and $8$) and for $n=4$ it
is $4$ (the values are $1$, $35/8$, $8$ and $16$). Unfortunately, these are all the values of $n$ for which
this simple code works, for larger $n$ another technique is needed.

Now I will show how the Fig.~\ref{fig:o} was obtained. The key point is the function
\begin{widetext}
\begin{equation}
\begin{split}
    \In \mathtt{d[n\_, m\_] :=}\ &\mathtt{Function[c, FromDigits[f[c, Subsets[Range[n], m]] /@\ V[n], 2]]} \\
    &\mathtt{/@\ V[Plus\ @@\ (Binomial[n, \#] \& /@\ Range[0, m])]}
\end{split}
\end{equation}
\end{widetext}
\noindent which returns the numbers $B_n(f)$ of the boolean functions $f \in F_n$ of degree $\leqslant m$. To
visualize boolean functions with respect to their degree let us create a list $\mathtt{dt = Table[n,
\{2^{2^n}\}]};$. Then we can fill it in as
\begin{equation}
    \mathtt{For[i=1, i \leqslant n, i++, dt [[ d[n, n-i]+1 ]] = n-i]}
\end{equation}
The table $\mathtt{Partition[dt, 2^{2^{n-1}}]}$ gives the desired visualization.

Now let us discuss the equivalence of Bell inequalities. The map $p_\pi$ \eqref{eq:p} can be coded as
\begin{widetext}
\begin{equation}
    \In \mathtt{p[pi\_]:=\ Function[x, \#[Permute[x,InversePermutation[pi]]]]\&}
\end{equation}
\end{widetext}
\noindent The expression $\mathtt{p[pi]}$ is a functional: $\mathtt{p[pi][f]}$ gives the function $p_\pi f$,
and $\mathtt{p[pi][f][\{x_1, \ldots, x_n\}]}$ gives its value $(p_\pi f)(\vec{x})$ at $\vec{x} = (x_1,
\ldots, x_n)$. The maps $\delta$, $s_{\vec{y}}$ \eqref{eq:deltas} and $t_{\vec{z}}$ \eqref{eq:tz} can be
coded as
\begin{widetext}
\begin{equation}
\begin{split}
    \In &\mathtt{\delta :=\ Function[x, 1-\#[x]]\&} \\
    \In &\mathtt{s[y\_] :=\ Function[x, \#[Function[z, Mod[z, 2]] /@\ (x + y)]]\&} \\
    \In &\mathtt{t[z\_] :=\ Function[x, Mod[\#[x] + x.z, 2]]\&}
\end{split}
\end{equation}
\end{widetext}

Let us illustrate the relations between the maps under study. As an example, consider the first relation
\eqref{eq:pst}. I show that this relation is valid by applying the maps from both sides to all boolean
functions $f \in F_n$ and comparing the results. To do it, we need operations which produce the lists of
values $\{(p_\pi s_{\vec{y}} f)(\vec{x})\}$ and $\{(s_{\vec{y}} p_\pi f)(\vec{x})\}$, $\vec{x} \in V_n$ given
the list $\{f(\vec{x})\}$, $\vec{x} \in V_n$ of values of $f \in F_n$. The code
\begin{widetext}
\begin{equation}
\begin{split}
    \In \mathtt{ps[n\_, pi\_, y\_]}\ &\mathtt{:= Composition[p[pi], s[y]][Function[x, Part[\#, FromDigits[x,2]+1]]] /@\ V[n] \&} \\
    \In \mathtt{sp[n\_, pi\_, y\_]}\ &\mathtt{:= Composition[s[y], p[pi]][Function[x, Part[\#, FromDigits[x,2]+1]]] /@\ V[n] \&}
\end{split}
\end{equation}
\end{widetext}
\noindent is a solution to this problem. The expression
\begin{widetext}
\begin{equation}
    \In \mathtt{ps[3, \{2,3,1\}, \{1,1,0\}] /@\ V[2^3] - sp[3, \{2,3,1\}, \{1,0,1\}] /@\ V[2^3] //Short}
\end{equation}
\end{widetext}
\noindent produces a list of zero-lists, as expected. This example clearly demonstrates that the first
relation \eqref{eq:pst} is valid (in the case of $n=3$).
\begin{equation}
\begin{split}
    &\mathtt{\{\{0,0,0,0,0,0,0,0\}, <<254>>,} \\
    &\mathtt{\{0,0,0,0,0,0,0,0\}\}}
\end{split}
\end{equation}

Given a number $0 \leqslant B < 2^{2^n}$ the function
\begin{widetext}
\begin{equation}
\begin{split}
    \In \mathtt{e[n\_, B\_]}&\ \mathtt{:=\ Replace[FromDigits[Map[\#, V[n]], 2] \&} \\
    &\mathtt{/@\ Function[f, Apply[Function[\{\varepsilon, y, pi, z\}, Composition[\varepsilon, t[z], p[pi], s[y]]} \\
    &\mathtt{[Function[x, Part[IntegerDigits[f, 2, 2^n], FromDigits[x, 2] + 1]]]], \#]\&} \\
    &\mathtt{/@\ Tuples[\{\{Identity, \delta\}, V[n], Permutations[Range[n]], V[n]\}]][B]} \\
    &\mathtt{//Sort //Split, \{(i\_) ..\} \to i, 1]}
 \end{split}
\end{equation}
\end{widetext}
\noindent produces the list of numbers which correspond to the functions, equivalent to the one corresponding
to $B$. With this functions it is easy to get the equivalence classes. Let us illustrate the general idea by
the case of $n=3$. We start with $B=0$: $\mathtt{e[3, 0] //Short}$ produces
\begin{equation}
    \mathtt{\{0, 15, 51, <<10>>, 204, 240, 255\}}
\end{equation}
It is the first class of equivalence (which contains $16$ elements). The smallest number which is not in this
class is $1$; $\mathtt{e[3, 1] //Short}$ produces
\begin{equation}
    \mathtt{\{1, 2, 4, <<122>>, 251, 253, 254\}}
\end{equation}
It is the second class of equivalence (which contains $128$ elements). The smallest number which is not in
either class found so far is $3$; $\mathtt{e[3, 3] //Short}$ produces
\begin{equation}
    \mathtt{\{3, 5, 10, <<42>>, 245, 250, 252\}}
\end{equation}
It is the third class of equivalence (which contains $48$ elements). The ext number to try is $6$:
$\mathtt{e[3, 6] //Short}$ produces
\begin{equation}
    \mathtt{\{6, 9, 18, <<42>>, 237, 246, 249\}}
\end{equation}
It is the fourth class of equivalence (which also contains $48$ elements). The next number is $23$;
$\mathtt{e[3, 23] //Short}$ produces
\begin{equation}
    \mathtt{\{23, 24, 36, <<10>>, 219, 231, 232\}}
\end{equation}
it is the fifth class of equivalence (which contains $16$ elements). Since $16+128+48+48+16 = 256$, we
exhausted all numbers (boolean functions) which means that in the case of $n=3$ there are $5$ classes of
equivalence, they are shown in Fig.~\ref{sfig:bell-b}. The same approach allows one to get
Fig.~\ref{sfig:bell-b2}.

The maximal quantum violations can be found as follows. Let us introduce the functions
\begin{equation}
\begin{split}
    \In &\mathtt{S[\varphi\_]\ :=\ \{Cos[\varphi], I\ Sin[\varphi]\}} \\
    \In &\mathtt{S[\varphi\_List]\ :=\ Apply[Times, \#]\&} \\
    &\mathtt{/@\ Tuples[S /@\ \varphi]}
\end{split}
\end{equation}
The maximal violation \eqref{eq:v} can now be coded as
\begin{equation}
\begin{split}
    \In &\mathtt{v[f\_, \varphi\_]\ :=\ Abs[((-1)^{f[\#]}\&} \\
    &\mathtt{/@\ V[Length[\varphi]]).S[\varphi]]} \\
    \In &\mathtt{v2[B\_, \varphi\_]\ :=\ v[itof[Length[\varphi], B], \varphi]}
\end{split}
\end{equation}
The quantity $\mathtt{v2[B, \varphi]}$ give the maximal quantum violation $v_f$ of the function $f$
corresponding to the number $B$. Let us illustrate the calculation of maximal quantum violation for Mermin
inequalities. The Mermin inequalities (whose coefficients are given by \eqref{eq:Wo} and \eqref{eq:We}) can
be coded as (using the standard package \textsl{Algebra`SymmetricPolynomials})
\begin{equation}
    \mathtt{m[x\_]\ :=\ Mod[SymmetricPolynomial[x, 2], 2]}
\end{equation}
The maximal quantum violation of Mermin inequalities can be found as
\begin{equation}
\begin{split}
    \In &\mathtt{FindMaximum[v[m, \{\varphi_1, \ldots, \varphi_n\}],} \\
    &\mathtt{\{\varphi_1, \varphi^0_1\}, \ldots, \{\varphi_n, \varphi^0_n\}]}
\end{split}
\end{equation}
For example, the code
\begin{equation}
    \mathtt{FindMaximum[v[m, \{\varphi_1, \varphi_2\}], \{\varphi_1, 1\}, \{\varphi_2, 1\}]}
\end{equation}
produces
\begin{equation}
    \mathtt{\{1.41421, \{\varphi_1 \to\ 0.785398, \varphi_2 \to\ 0.785398\}\}}
\end{equation}
the code
\begin{equation}
\begin{split}
    &\mathtt{FindMaximum[v[m, \{\varphi_1, \varphi_2, \varphi_3\}],} \\
    &\mathtt{\{\varphi_1, 1\}, \{\varphi_2, 1\}, \{\varphi_3, 1\}]}
\end{split}
\end{equation}
produces
\begin{equation}
    \mathtt{\{2., \{\varphi_1 \to\ 0.785398, \varphi_2 \to\ 0.785398, \varphi_3 \to\ \ldots\}\}}
\end{equation}
in full agreement with the relation \eqref{eq:vm}. Since $v_f$ is the same for equivalent functions, using
the classes of equivalence calculated before, it is easy to get Fig.~\ref{fig:v}.

Now I will show how the table \ref{tbl:1} was obtained. To calculate the cycle index $Z_{\mathcal{J}_n}(x_1,
\ldots, x_{2^n})$ we need to calculate the sum \eqref{eq:Z}. The vectors $\vec{c}$ for a given $n$ can be
obtained using the standard package $\textit{Combinatorica}$, containing the function
$\mathtt{Partitions[n]}$, which returns the list of partitions of $n$. Any partition $p$ of this list has the
following form
\begin{equation}
    p = \{\underbrace{n_1, \ldots, n_1}_{k_1}, \underbrace{n_2, \ldots, n_2}_{k_2}, n_3, \ldots\},
\end{equation}
with $n_1 > n_2 > n_3 > \ldots$ and $k_1 n_1 + k_2 n_2 + \ldots = n$. The relation
\begin{equation}
    p \to \vec{c} = \{\underbrace{\overbrace{\ldots, k_2}^{n_2}, \ldots, k_1}_{n_1}, \ldots\}
\end{equation}
gives a one-to-one correspondence between the partitions of $n$ and the vectors $\vec{c}$. Here $k_i$ is on
$n_i$-th place and the other components are zero. This correspondence can be realized in \Mathematica via
\begin{widetext}
\begin{equation}
\begin{split}
    \In \mathtt{ct[p\_] :=}\ &\mathtt{Module[\{c=Table[0, \{Plus\ @@\ p\}]\},} \\
    &\mspace{20mu} \mathtt{Set[Part[c, \#[[1]]], Length[\#]]\& /@\ Split[p];} \\
    &\mspace{20mu} \mathtt{Return[c]} \\
    &\mathtt{]}
\end{split}
\end{equation}
\end{widetext}
\noindent Then we need to define the cross-product. We introduce the object $\mathtt{var[p, n]}$ which
represents the $n$-th power of the $p$-th independent variable. The code
\begin{widetext}
\begin{equation}
    \In \mathtt{cp[var[p\_, n\_], var[q\_, m\_]]\ :=\ var[LCM[p, q], n\ m\ GCD[p, q]]}
\end{equation}
\end{widetext}
\noindent reproduces the definition \eqref{eq:x}. To extend the cross-product for arbitrary polynomials we
need the following definitions (the order in which they are given is important):
\begin{widetext}
\begin{equation}
\begin{split}
    \In &\mathtt{cp[v\_, c\_]          := c\ v /;\ NumericQ[c]} \\
    \In &\mathtt{cp[c\_, v\_]          := c\ v /;\ NumericQ[c]} \\
    \In &\mathtt{cp[v1\_ + v2\_, v3\_] := cp[v1, v3] + cp[v2, v3]} \\
    \In &\mathtt{cp[v3\_, v1\_ + v2\_] := cp[v3, v1] + cp[v3, v2]} \\
    \In &\mathtt{cp[v3\_, v1\_ v2\_]   := cp[v3, v1]\ cp[v3, v2]} \\
    \In &\mathtt{cp[v1\_ v2\_, v3\_]   := cp[v1, v3]\ cp[v2, v3]} \\
    \In &\mathtt{cp[c\_ v1\_, v2\_]    := c\ cp[v1, v2] /;\ NumericQ[c]} \\
    \In &\mathtt{cp[v1\_, c\_ v2\_]    := c\ cp[v1, v2] /;\ NumericQ[c]}
\end{split}
\end{equation}
\end{widetext}
\noindent The following code reproduces the identities $(x^n_p)^m = x^{n m}_p$ and $x^n_p x^m_p = x^{n+m}_p$
respectively:
\begin{widetext}
\begin{equation}
\begin{split}
    \In &\mathtt{var /:\ Power[var[p\_, n\_], m\_]\ :=\ var[p, n\ m]} \\
    \In &\mathtt{var /:\ var[p\_, n\_]\ var[p\_, m\_]\ :=\ var[p, n + m]}
\end{split}
\end{equation}
\end{widetext}
The code for the functions $a$ \eqref{eq:a} and $b$ \eqref{eq:b} is obvious:
\begin{widetext}
\begin{equation}
\begin{split}
    \In &\mathtt{a[k\_]\ :=\ \frac{1}{k} Plus\ @@\ \left(2^\#\ MoebiusMu\left[\frac{k}{\#}\right]\& /@\ Divisors[k]\right)} \\
    \In &\mathtt{b[k\_]\ :=\ \frac{1}{k} Plus\ @@\ \left(2^{\#/2} MoebiusMu\left[\frac{k}{\#}\right]\&
    /@\ Complement[Divisors[k], \ Divisors[k/2]]\right)}
\end{split}
\end{equation}
\end{widetext}
\noindent A factor of of the sum \eqref{eq:Z} can be coded as
\begin{widetext}
\begin{equation}
\begin{split}
    \In \mathtt{factor[i\_]\ :=}\ &\mathtt{Times\ @@\ (var[\#, a[\#]] \& /@\ Divisors[i]) +
    Times\ @@\ (var[\#, b[\#]] \&} \\
    &\mathtt{/@\ Complement[Divisors[2i], Divisors[i]])}
\end{split}
\end{equation}
\end{widetext}
\noindent A term of the sum is coded as
\begin{widetext}
\begin{equation}
\begin{split}
    \In &\mathtt{term[c\_]\ :=\ Module[\{t\},} \\
    &\mspace{20mu}\mathtt{t = MapIndexed[If[\#1 == 0, 1, If[\#1 == 1, factor[\#2[[1]]],} \\
    &\mspace{40mu}\mathtt{Fold[cp, factor[\#2[[1]]], Table[factor[\#2[[1]]],\{\#1-1\}]]]]\&, c];} \\
    &\mspace{20mu}\mathtt{Fold[cp, First[t], Rest[t]]} \\
    &\mathtt{] /(Times\ @@\ (Factorial /@\ c) Times\ @@\ MapIndexed[(2 \#2[[1]])^{\#1} \&, c])}
\end{split}
\end{equation}
\end{widetext}
\noindent The whole sum \eqref{eq:Z} is given by
\begin{widetext}
\begin{equation}
    \In \mathtt{cycleIndex[n\_]\ :=\ Plus\ @@\ (term /@\ (ct /@\ Partitions[n]))}
\end{equation}
\end{widetext}
The number $\overline{N}_n$ can be obtained as follows:
\begin{widetext}
\begin{equation}
\begin{split}
    \In \mathtt{ub[n\_]\ :=}\ &\mathtt{\frac{1}{2} ((cycleIndex[n]\ /.\ \{var[p\_, k\_]\ \to\ 2^k\})} \\
          &\mathtt{+ (cycleIndex[n]\ /.\ \{var[p\_, k\_]\ :\to\ If[EvenQ[p], 2^k, 0]\}))}
\end{split}
\end{equation}
\end{widetext}
Then one can get the numbers form the table \ref{tbl:1}: the code $\mathtt{Table[ub[n], \{n, 1, 5\}]}$
produces $\mathtt{\{2, 4, 14, 222, 616126\}}$.

\section{Fast Walsch-Hadamard transform}

A much faster way of calculating the Walsch-Hadamar transform of a boolean function $f \in F_n$ (i.e.
calculation of all $2^n$ numbers $W_f(\vec{u})$, $\vec{u} \in V_n$) is based on the following decomposition
of the Hadamard matrix:
\begin{equation}
    H_n = \prod^n_{k=1} M_{n-k+1} \equiv \prod^n_{k=1} (E_{n-k} \otimes H \otimes E_{k-1}),
\end{equation}
which can be proved by induction. The relation $\vec{y} = M_k \vec{x}$ can be written as
\begin{equation}
\begin{split}
    y_{j 2^k + i} &= x_{j 2^k + i} + x_{j 2^k + 2^{k-1} + i}, \\
    y_{j 2^k + 2^{k-1} + i} &= x_{j 2^k + i} - x_{j 2^k + 2^{k-1} + i},
\end{split}
\end{equation}
for $i = 1, \ldots, 2^{k-1}$ and $j = 0, \ldots, 2^{n-k}-1$. The Walsh-Hadamard transform can be calculated
according to \eqref{eq:WH2} as
\begin{equation}
    \vec{w}_f = M_n \ldots M_1 \vec{z}_f.
\end{equation}
Starting with the vector $\vec{x} = \vec{z}_f$ we calculate vectors $\vec{y}_k = M_k \ldots M_1 \vec{z}_f$
for $k = 1, \ldots, n$. Then $\vec{y}_n$ is $\vec{w}_f$. The algorithm in \textsl{C} is presented in the
function \textsl{whtl} in the file \textsl{whtl.c} below. To turn \textsl{whtl.c} into a program usable from
inside \Mathematica another file, \textsl{whtl.tm} is necessary.

\begin{Verbatim}[frame=lines, framesep=3mm, labelposition=topline, label={\normalfont whtl.tm}]
 :Begin:
 :Function:            whtl
 :Pattern:              WHTL[x_List]
 :Arguments:          {x}
 :ArgumentTypes:   {IntegerList}
 :ReturnType:         Manual
 :End:
\end{Verbatim}

\begin{Verbatim}[frame=lines, framesep=3mm, labelposition=topline,
commandchars=\\\{\}, codes={\catcode`$=3\catcode`^=7}, label={\normalfont whtl.c}]
 \textcolor{blue}{#include} "mathlink.h"
 \textcolor{blue}{#include} <stdlib.h>
 \textcolor{blue}{#include} <string.h>

 \textcolor{gray}{// \textit{xlen is of the form $2^n$, returns $n$}}
 \textcolor{blue}{long} __log(\textcolor{blue}{long} xlen) \{
   \textcolor{blue}{long} i = 1, m = xlen;
   \textcolor{blue}{while}((m >>= \textcolor{red}{1}) != \textcolor{red}{1}) i++;
   \textcolor{blue}{return} i;
 \}

 \textcolor{blue}{void} whtl(\textcolor{blue}{int *}x, \textcolor{blue}{long} xlen) \{
   \textcolor{blue}{int *}y = (\textcolor{blue}{int*})calloc(xlen, \textcolor{blue}{sizeof}(\textcolor{blue}{int}));
   \textcolor{blue}{long} i, j, k, t, pow1, pow2, ind1, ind2;
   \textcolor{blue}{long} n = __log(xlen);

   \textcolor{blue}{for}(k = \textcolor{red}{1}; k <= n; k++) \{
     pow1 = \textcolor{red}{1}<<(k-\textcolor{red}{1});
     pow2 = \textcolor{red}{1}<<(n-k);
     \textcolor{blue}{for}(j = \textcolor{red}{0}; j < pow2; j++) \{
       t = j<<k;
       \textcolor{gray}{// \textit{in C arrays are numbered from 0}}
       \textcolor{blue}{for}(i = \textcolor{red}{0}; i < pow1; i++) \{
         ind1 = t + i;
         ind2 = ind1 + pow1;
         y[ind1] = x[ind1] + x[ind2];
         y[ind2] = x[ind1] - x[ind2];
       \}
     \}
     \textcolor{blue}{if}(k < n)
       memcpy(x, y, \textcolor{blue}{sizeof}(\textcolor{blue}{int})*xlen);
   \}

   MLPutIntegerList(stdlink, y, xlen);
   free(y);
 \}

 \textcolor{blue}{int} main(\textcolor{blue}{int} argc, \textcolor{blue}{char*} argv[]) \{
   \textcolor{blue}{return} MLMain(argc, argv);
 \}
\end{Verbatim}

On a UNIX system the files are compiled with the following command
\begin{equation}
    \mathtt{\$\ mcc\ -o\ WHTL\ whtl.tm\ whtl.c}
\end{equation}
This command produces an executable file \textsl{WHTL}. To use it in \Mathematica it must be installed as
\begin{equation}
    \mathtt{Install["/path/to/program/WHTL"];}
\end{equation}
Then it can be used as
\begin{equation}
    \mathtt{WHTL[\{z_1, \ldots, z_{2^n}\}]}
\end{equation}
where all $z_i$, $i = 1, \ldots, 2^n$ are $\pm 1$.

\end{document}